\documentclass[twocolumn,twocolappendix,tighten]{aastex7}

\usepackage{soul}
\usepackage{array}
\usepackage{tabularx}
\usepackage{float}

\begin{document}

\title{SN 2024abfl: A Flat-Plateau, Low-Luminosity Type IIP Supernova with Early CSM Interaction}

\newcommand{\UTA}{\affiliation{Department of Astronomy, The University of Texas at Austin, 2515 Speedway, Austin, TX 78712-1205, USA}}

\newcommand{\MMA}{\affiliation{Maria Mitchell Association, 4 Vestal St., Nantucket, MA 02554, USA}}

\newcommand{\UA}{\affiliation{Steward Observatory, University of Arizona, 933 North Cherry Avenue, Tucson, AZ 85721-0065, USA}}

\newcommand{\GNL}{\affiliation{Gemini Observatory/NSF's NOIRLab, 670 N. A'ohoku Place, Hilo, HI 96720, USA}}

\newcommand{\SSI}{\affiliation{Space Science Institute, 4765 Walnut St., Suite B, Boulder, CO 80301, USA}}

\newcommand{\UW}{\affiliation {DIRAC Institute, Department of Astronomy, University of Washington, 3910 15th Avenue NE, Seattle, WA 98195, USA}}

\newcommand{\keck}{\affiliation{W.M. Keck Observatory, 65-1120 Mamalahoa Highway, Kamuela, HI 96743, USA}}

\newcommand{\UCSD}{\affiliation{Department of Astronomy \& Astrophysics, University of California, San Diego, 9500 Gilman Drive, MC 0424, La Jolla, CA 92093-0424, USA}}

\newcommand{\LCO}{\affiliation{Las Cumbres Observatory, 6740 Cortona Drive, Suite 102, Goleta, CA 93117-5575, USA}}

\newcommand{\UCSB}{\affiliation{Department of Physics, University of California, Santa Barbara, CA 93106-9530, USA}}

\newcommand{\KITP}{\affiliation{Kavli Institute for Theoretical Physics, University of California, Santa Barbara, CA 93106-4030, USA}}

\newcommand{\UCD}{\affiliation{Department of Physics and Astronomy, University of California, Davis, 1 Shields Avenue, Davis, CA 95616-5270, USA}}

\newcommand{\CfA}{\affiliation{Center for Astrophysics \textbar{} Harvard \& Smithsonian, 60 Garden Street, Cambridge, MA 02138-1516, USA}}

\newcommand{\IAIFI}{\affiliation{The NSF AI Institute for Artificial Intelligence and Fundamental Interactions}}

\newcommand{\USask}{\affiliation{Department of Physics \& Engineering Physics, University of Saskatchewan, 116 Science Place, Saskatoon, SK S7N 5E2, Canada}}

\newcommand{\Rut}{\affiliation{Department of Physics and Astronomy, Rutgers, the State University of New Jersey,136 Frelinghuysen Road, Piscataway, NJ 08854-8019, USA}}

\newcommand{\JHU}{\affiliation{Department of Physics and Astronomy, The Johns Hopkins University, 3400 North Charles Street, Baltimore, MD 21218, USA}}

\newcommand{\TT}{\affiliation{Department of Physics \& Astronomy, Texas Tech University, Lubbock, TX 79410-1051, USA}}

\newcommand{\LSU}{\affiliation{Department of Physics \& Astronomy, Louisiana State University, Baton Rouge, LA 70803, USA}}

\newcommand{\IA}{\affiliation{Istituto Nazionale di Astrofisica, Osservatorio Astronomico di Brera, via E. Bianchi 46, 23807 Merate (LC), Italy}}

\newcommand{\MAS}{\affiliation{Millennium Institute of Astrophysics (MAS), Nuncio Monsenor Sòtero Sanz 100, Providencia, Santiago RM, Chile}}

\newcommand{\Catalyst}{\altaffiliation{LSSTC Catalyst Fellow}}

\newcommand{\Monash}{\affiliation{School of Physics and Astronomy, Monash University, Clayton, Australia}}

\newcommand{\OzGrav}{\affiliation{OzGrav: The ARC Center of Excellence for Gravitational Wave Discovery, Australia}}

\newcommand{\CIERA}{\affiliation{Center for Interdisciplinary Exploration and Research in Astrophysics (CIERA), 1800 Sherman Ave., Evanston, IL 60201, USA}}

\newcommand{\NU}{\affiliation{Department of Physics and Astronomy, Northwestern University, 2145 Sheridan Road, Evanston, IL 60208, USA}}

\newcommand{\Konkoly}{\affiliation{Konkoly Observatory, HUN-REN Research Center for Astronomy and Earth Sciences, Konkoly Th. M. út 15-17., Budapest, 1121 Hungary; MTA Center of Excellence}}

\newcommand{\Szeged}{\affiliation{Department of Experimental Physics, Institute of Physics, University of Szeged, D\'om t\'er 9, Szeged, 6720 Hungary}}

\newcommand{\Tsinghua}{\affiliation{Physics Department, Tsinghua University, Beijing 100084, People’s Republic of China}}

\newcommand{\UF}{\affiliation{Department of Astronomy, University of Florida, 211 Bryant Space Science Center, Gainesville, FL 32611-2055, USA}}
\author[0009-0007-3225-2543]{Madison Gerard}
\MMA
\UTA
\email{maddiekgerard@gmail.com}

\author[0000-0003-0123-0062]{Jennifer E. Andrews}
\GNL
\email{Jennifer.Andrews@noirlab.edu}

\author[0000-0002-0141-7436]{Geoffrey C. Clayton}
\MMA
\SSI
\LSU
\email{gclayton@spacescience.org}

\author[0000-0003-4102-380X]{David J. Sand}
\UA
\email{dsand@arizona.edu}

\author[0000-0002-4924-444X]{K. Azalee Bostroem}
\UA\Catalyst\email{bostroem@arizona.edu}

\author[0000-0002-0744-0047]{Jeniveve Pearson}
\UA\email{jenivevepearson@arizona.edu}

\author[0000-0001-6191-7160]{Raya Dastidar}
\IA\email{rdastidr@gmail.com}

\author[0000-0002-7352-7845]{Aravind P. Ravi}
\UCD\email{apazhayathravi@ucdavis.edu}

\author[0000-0003-4175-4960]{Conor L. Ransome}
\UA\email{cransome@arizona.edu}

\author[0000-0001-8073-8731]{Bhagya Subrayan}
\UA\email{bsubrayan@arizona.edu}

\author[0000-0002-0832-2974]{Griffin Hosseinzadeh}
\UCSD\email{ghosseinzadeh@ucsd.edu}

\author[0000-0002-9454-1742]{Brian Hsu}
\UA\email{bhsu@arizona.edu}

\author[0000-0002-7937-6371]{Yize Dong} 
\CfA
\email{yize.dong@cfa.harvard.edu}

\author[orcid=0000-0002-4022-1874, gname=Manisha, sname=Shrestha]{Manisha Shrestha}
\Monash\OzGrav\email{manisha.shrestha@monash.edu}

\author[0000-0001-8818-0795]{Stefano Valenti}
\UCD\email{}

\author[orcid=0000-0001-5510-2424, gname=Nathan, sname=Smith]{Nathan Smith}
\UA \email{nathansmith@arizona.edu}

\author[0000-0003-0549-3281]{Daryl Janzen}
\USask \email{}

\author[0000-0001-9589-3793]{M.~J. Lundquist}
\keck \email{}

\author[0000-0002-7015-3446]{Nicol\'as Meza}
\UCD \email{}

\author[0000-0001-8738-6011]{Saurabh W.\ Jha}
\Rut \email{}

\author[orcid=0000-0002-8297-2473, gname=Kate, sname=Alexander]{Kate D. Alexander}
\UA \email{kdalexander@arizona.edu}

\author[orcid=0000-0003-0528-202X, gname=Collin, sname=Christy]{Collin Christy}
\UA \email{collinchristy@arizona.edu}

\author[orcid=0000-0003-4537-3575, gname=Noah, sname=Franz]{Noah Franz}
\UA \email{nfranz@arizona.edu}

\author[orcid=0000-0003-3108-1328, gname=Lindsey, sname=Kwok]{Lindsey A. Kwok}
\CIERA \NU \email{lindsey.kwok@northwestern.edu}

\author[0000-0002-1895-6639]{Moira Andrews}
\LCO \UCSB \email{}

\author[0000-0003-4914-5625]{Joseph Farah}
\LCO \UCSB \email{}

\author[0000-0002-1125-9187]{Daichi Hiramatsu}
\UF\email{}

\author[0000-0003-4253-656X]{D.\ Andrew Howell}
\LCO \UCSB \email{}

\author[0000-0001-5807-7893]{Curtis McCully}
\LCO \email{}

\author[orcid=0009-0006-7296-728X]{Kathryn Wynn}
\LCO \UCSB \email{}

\author[0000-0002-8770-6764]{R\'eka K\"onyves-T\'oth}
\Konkoly\Szeged\email{konyvestoth.reka@csfk.org}

\author[orcid=0000-0002-7334-2357, gname=Xiaofeng, sname=Wang]{Xiaofeng Wang} 
\Tsinghua \email{wang_xf@mail.tsinghua.edu.cn}

\received{}
\revised{}
\accepted{}
\submitjournal{\textit{The Astrophysical Journal}}

\begin{abstract}
We present photometric and spectroscopic observations of SN 2024abfl, a low-luminosity Type IIP supernova (LLSN) discovered shortly after explosion. The transient reached a peak absolute magnitude of $M_V = -14.9$ and exhibited an extended, flat plateau lasting $\sim$125 days. From the late-time bolometric light curve, we estimate a $^{56}$Ni mass of $\sim0.01~M_\odot$, consistent with other LLSNe. Analytical shock-cooling models fail to reproduce the rapid early rise, indicating that circumstellar material (CSM) interaction contributed to the initial emission. The spectroscopic evolution is typical of LLSNe, with relatively narrow metal lines and low expansion velocities ($\lesssim 3000$ km s$^{-1}$) that decline slowly over time. We detect a broad “ledge” feature around 4600 \AA~within three days of explosion, which we interpret as a blend of high-ionization shock-accelerated CSM lines. Multi-peaked H$\alpha$ profiles develop during the plateau phase, consistent with complex ejecta–CSM interactions. As one of the best-observed examples of LLSNe, SN 2024abfl exhibits a weak explosion and signatures of nearby CSM, offering new insights into progenitor properties, pre-explosion mass loss, and the diversity of LLSNe.
\end{abstract}

\keywords{Circumstellar matter (241), Core-collapse supernovae (304), Red supergiant stars (1375), Supernovae (1668), Type II supernovae (1731)}

\section{Introduction}
Type II supernovae (SNe II) are produced by the core-collapse (CC) of massive stars ($\gtrsim 8~M_\odot$) and are characterized by the presence of hydrogen in their spectra. SNe II are broadly divided into two subcategories, IIP and IIL, based on the shape of their light curves \citep{1979A&A....72..287B, 1997ARA&A..35..309F, 2009ARA&A..47...63S}, although current observational evidence may suggest a continuous class of objects \citep{2014ApJ...786...67A, 2016MNRAS.459.3939V, 2021ApJ...913...55H}. Type IIL SNe exhibit a nearly linear decline in brightness following peak, whereas Type IIP SNe exhibit an extended plateau phase, typically lasting $\sim$100 days, followed by a decline dominated by radioactive decay \citep{1979A&A....72..287B}. 

Recent studies have found the photometric and spectral properties of SNe II to be diverse, with peak \textit{V-}band luminosities ranging from $-19 < M_V < -13$ \citep{2014ApJ...786...67A, 2015ApJ...799..208S, 2016MNRAS.455.4087G, 2016MNRAS.459.3939V, 2019MNRAS.490.2799D}. The extrema of the SNe II distribution have been the focus of intense study, and SNe II with peak magnitudes $M_V \geq -15.5$ are classified as low-luminosity SNe II \citep[LLSNe;][]{2004MNRAS.347...74P}. These faint events are relatively rare, comprising only $5{-}8\%$ of the total SNe II population \citep{2004MNRAS.347...74P, 2025PASP..137d4203D}. Given their rarity, each new LLSN provides an important opportunity to test how explosion energy, nickel yield, and progenitor properties vary at the faint end of the SNe II population.

Compared to the more typical SNe II, LLSNe exhibit slower expansion velocities (a few $10^3$ km s$^{-1}$), lower explosion energies ($\leq$ a few times $10^{50}$ erg), and lower nickel masses \citep[$\leq 10^{-2}M_\odot$;][]{1998ApJ...498L.129T, 2004MNRAS.347...74P, 2014MNRAS.439.2873S, 2017MNRAS.464.3013P}. In addition, their light curves feature extended plateau phases relative to standard SNe II, often lasting $\sim$100–130 days.

The prototype of this class is SN 1997D \citep{1998ApJ...498L.129T, 2001MNRAS.322..361B}, which, at the time, was the faintest SN II ever discovered, peaking at $M_V \lesssim -14.5$ mag. Two main scenarios have been proposed to explain the characteristics of SN 1997D. The first involves a low-energy explosion of a high-mass star ($\gtrsim 25~M_\odot$), in which a substantial portion of stellar material falls back onto the newly formed compact remnant \citep{1998ApJ...498L.129T, 1998ApJ...505..876Z}. The alternative scenario suggests the core collapse of a star with an initial ZAMS mass of $8\text{–}10~M_\odot$, which would undergo extensive post–main-sequence evolution before exploding as a super asymptotic giant branch (sAGB) star or low-mass red supergiant (RSG) \citep{2000A&A...354..557C}.

In recent years, hydrodynamical models \citep{2013MNRAS.433.1745D, 2014MNRAS.439.2873S, 2019A&A...629A.124M, 2022MNRAS.514.4173K} and constraints from archival pre-explosion Hubble Space Telescope (HST) images, such as those of SN 2002gd \citep{2003PASP..115....1V}, SN 2005cs \citep{2005MNRAS.360..288M, 2006ApJ...641.1060L}, and SN 2008bk \citep{2012AJ....143...19V}, have found LLSNe to arise from low-mass ($\lesssim 12~M_\odot$) RSGs. Additionally, electron-capture supernovae (ECSNe)—a subclass of core-collapse supernovae resulting from the collapse of a degenerate ONeMg core in sAGB stars—have also been suggested as a potential origin for some LLSNe \citep{2018ApJ...861...63H, 2021NatAs...5..903H, 2022MNRAS.513.4983V}. Reliably distinguishing ECSNe from low-luminosity iron core-collapse SNe (LL-CCSNe) remains challenging, as some models predict that ECSNe can appear nearly identical to LL-CCSNe, with progenitors occupying a similar initial mass range \citep[$8\text{--}10~M_\odot$;][]{2006A&A...450..345K} to low-mass RSGs that undergo iron core collapse \citep{1984ApJ...277..791N}.

If a SN is observed within a few days of explosion, narrow emission lines may be present, providing insight into the progenitor system. ECSNe, originating from sAGB stars, are expected to be surrounded by denser circumstellar material (CSM) than low-mass RSGs, owing to their higher mass-loss rates at comparable luminosities during the late stages of stellar evolution, making early-time spectroscopy a potentially powerful diagnostic. These narrow lines, often referred to as “flash” spectroscopy, reveal the composition, density, and velocity of the CSM \citep{2014Natur.509..471G}. The most common interpretation is that they arise from recombination of CSM ionized by the shock-breakout flash or very early ejecta–CSM interaction \citep{2015MNRAS.449.1876S, 2016ApJ...818....3K}. Observations show that flash-ionized features occur in $\sim30\%$ SNe II \citep{2021ApJ...912...46B}, highlighting the importance of early-time observations for probing progenitor mass-loss histories.

Rather than the narrow features, some SNe II show broad, blueshifted spectral features near 4600 $\text{\AA}$ in the days following explosion \citep{2020ApJ...902....6S}. These broad features, known as a `ledge', have been observed in several faint SNe II, including SN 2005cs \citep{2006MNRAS.370.1752P, 2009MNRAS.394.2266P}, 2016bkv \citep{2018ApJ...861...63H}, 2018lab \citep{2023ApJ...945..107P}, 2021gmj \citep{2024ApJ...971..141M}, and 2022acko \citep{2023ApJ...953L..18B, 2025MNRAS.540.2591L}. The feature has also been seen in normal Type IIP SNe \citep{2007ApJ...666.1093Q, 2018MNRAS.476.1497B, 2019ApJ...885...43A}. Though interpretations of the `ledge' feature vary, it could indicate an RSG progenitor for LLSNe, rather than an sAGB, with an extended envelope exploding into low-density CSM \citep{2022ApJ...935...31H, 2023ApJ...945..107P}.

In this work, we present an analysis of the photometric and spectroscopic characteristics of the low-luminosity SN 2024abfl. SN 2024abfl was discovered early, allowing for a detailed study of its early-time `ledge' feature and subsequent photometric evolution. In Section \ref{sec:disc}, we review the discovery and classification of SN 2024abfl and report the properties of the SN and its host galaxy, NGC 2146. Section \ref{sec:obs} describes our observations and data-reduction procedures, while Sections \ref{sec:phot} and \ref{sec:spec} present the photometric and spectroscopic evolution of SN 2024abfl, respectively. Finally, our conclusions are summarized in Section \ref{sec:summary}.

\begin{figure}
    \centering
    \includegraphics[width=\columnwidth]{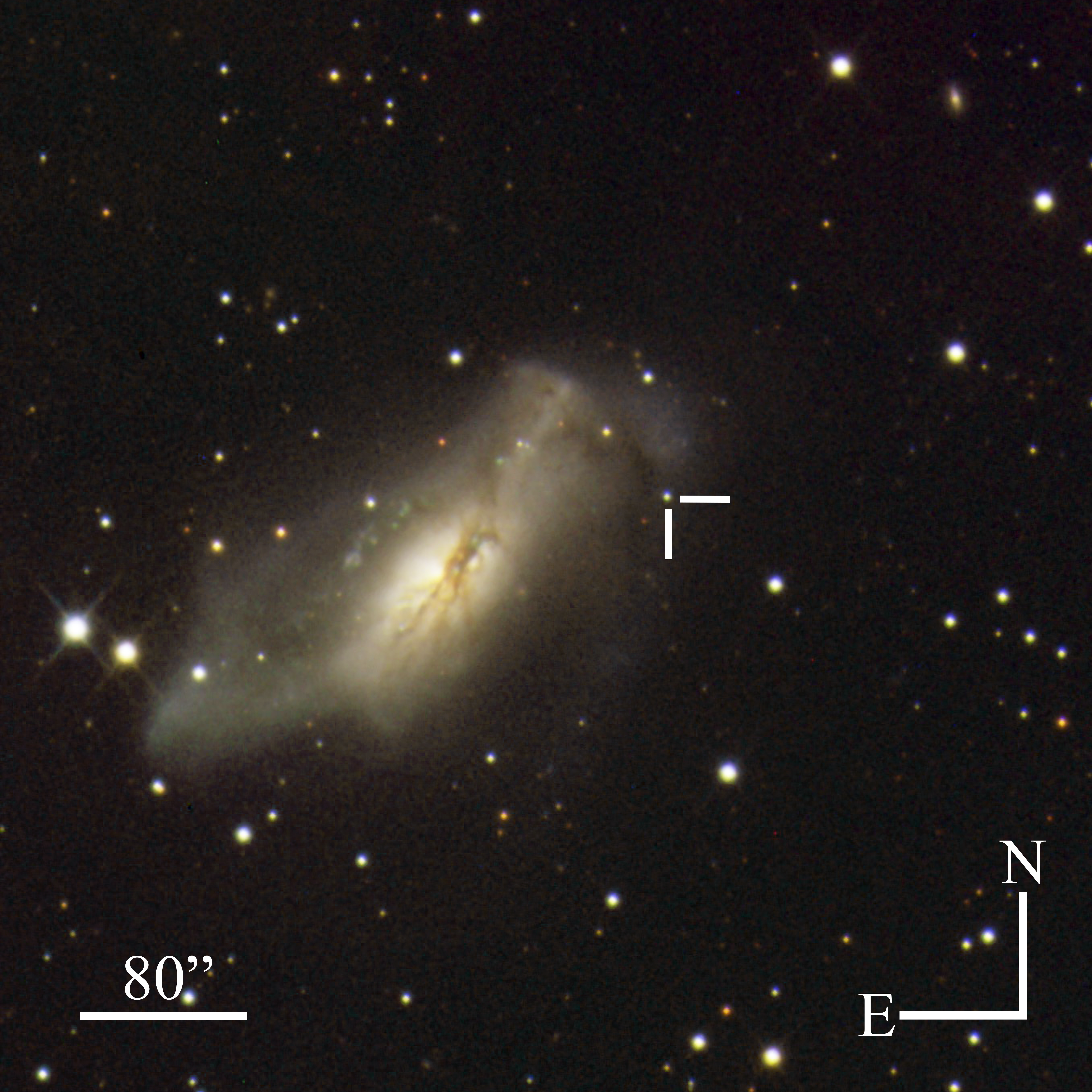}
    \caption{Composite $g,r,i$ image of SN 2024abfl in NGC 2146 obtained by Las Cumbres Observatory on 2024 December 4. The SN location is marked with white crosshairs.}
    \label{fig:pic}
\end{figure}

\section{Discovery, Distance, and Reddening}
\label{sec:disc}
\subsection{Discovery and Classification}
SN 2024abfl was discovered in NGC 2146 (Figure \ref{fig:pic}) at J2000 coordinates $\alpha = 06^{\mathrm{h}},18^{\mathrm{m}},01\fs140$, $\delta = +78^\circ,22',01\farcs52$, on 2024 November 15 at 14:00:57 UTC \citep[MJD~60629.58;][]{2024TNSTR4506....1I}. The SN was discovered at 17.5 mag in the clear filter ($M_{\mathrm{clear}} \approx -14$) and was not detected in the same filter down to 19 mag two days earlier \citep[]{2024TNSTR4506....1I}. The redshift of SN~2024abfl is $z = 0.002979$, measured from H$\alpha$ in the first spectrum obtained 2.6 days after explosion, and is adopted throughout this work.

SN 2024abfl was independently observed by the Zwicky Transient Facility \citep[ZTF;][]{2019PASP..131a8002B, 2019PASP..131g8001G} on 2024 November 14 at 06:42:26 UTC (MJD 60628.28). The automated pipeline did not detect  the source down to a limiting magnitude of $r = 19.7$ ($M_r \approx -11.0$), and this observation was formally classified as a nondetection. However, visual inspection clearly revealed the presence of the source, and forced PSF photometry recovered a significant detection of $r = 17.4 \pm 0.02$ ($M_r \approx -14.1$). 
Similarly, ATLAS \citep{2018PASP..130f4505T} initially reported only upper limits on 2024 November 15 at 13:28:26 UTC, but we performed forced photometry \citep{2020PASP..132h5002S} via the SAGUARO TOM \citep{2024ApJ...964...35H} and measured the SN at $o = 17.18 \pm 0.13$. Based on these observations, we adopt an explosion epoch of MJD 60627.91, corresponding to the midpoint between the last non-detection (November 13) and the first detection (November 14).

Spectroscopic follow-up performed on 2024 November 15 with the 2m Himalayan Chandra Telescope classified the object as a Type II SN, showing similarities to the Type IIP SN 1999gi \citep{2024TNSCR4515....1D}. Shortly after, we obtained a Gemini-North spectrum of SN 2024abfl using our rapid response long-term program (GN-24B-LP-112; PI Sand $\&$ Andrews), confirming the object as a Type II SN \citep{2024TNSCR4535....1A}. 

\subsection{Archival HST data}\label{sec:hst}
Fortuitously, SN 2018zd exploded just 6\arcsec\ away from the position of SN 2024abfl in NGC 2146. SN~2018zd is one of the first viable candidates for an ECSN, a Type II supernova that occurs when the core of an sAGB star undergoes electron capture, leading to gravitational collapse into a neutron star \citep{1980PASJ...32..303M, 1984ApJ...277..791N}. Due to the peculiarity of SN 2018zd, HST has targeted this region of NGC 2146 numerous times since 2018. All of these observations included the position of SN 2024abfl. From the archival data, \citet{2025ApJ...982L..55L} found a star-like object at the location of SN 2024abfl with an average F814W magnitude of $m_{F814W} \sim 25.2$ mag. The color and brightness of this source are consistent with a moderately reddened RSG, leading to its identification as a potential progenitor of SN 2024abfl, with an estimated mass of $\sim 9$–12~$M_\odot$. 

\subsection{Distance and Reddening}
The exact distance to the host galaxy NGC 2146 has been the subject of considerable debate in the analyses of SN 2018zd \citep{2021arXiv210912943C, 2021NatAs...5..903H}. This uncertainty arises because NGC 2146 lies at an intermediate distance, which complicates distance measurements. The galaxy is too close for redshift-based distances to be reliable given potential peculiar velocities, yet it is too distant for precise TRGB- or Cepheid-based distances to be available without space-based observations.

We adopt a distance estimate of $15.6^{+6.1}_{-3.0}$~Mpc, based on the combined probability distribution from kinematic, Tully–Fisher, and globular cluster size measurements presented by \citet{2021arXiv210912943C}. This range corresponds to the 16th and 84th percentiles of the summed probability density function and corresponds to a distance modulus of $\mu = 30.97^{+0.71}_{-0.47}$~mag. Note that the largest distance encompassed by the uncertainty would technically increase the peak magnitude of SN~2024abfl to above that of a LLSN. 

As an independent check, we also estimate the distance using the SN itself via the `standardized candle method' \citep[SCM;][]{2002ApJ...566L..63H, 2006ApJ...645..841N} for SNe IIP. The SCM uses the I-band apparent brightness and the expansion velocity of the \ion{Fe}{2} $\lambda 5169$ line at day 50, approximately the midpoint of the SN II plateau, denoted as $I_{50}$ and $v_{50}$, respectively. We follow the calibration of \citet{2015A&A...580L..15P}, based on several SNe II-P in hosts with Cepheid distances.

Using the calibration from \citet{2005AJ....130..873J}, we convert our observed i-band magnitude on day 49.9 to the Johnson–Cousins I system required for the SCM, obtaining $I_{50} = 16.34 \pm 0.10$ mag. From the spectrum at day 49, we measure $v_{50} = 1500 \pm 23.7$ km s$^{-1}$, assuming minimal evolution to day 50. Applying the \citet{2015A&A...580L..15P} relation, we derive an SCM distance of $14.42 \pm 1.87$ Mpc ($\mu = 30.79 \pm 0.28$ mag), consistent with the Tully–Fisher-based estimate. 
The SCM distance provides a useful reference for verifying that our photometric and velocity calibrations are consistent with established extragalactic distance scales. Nevertheless, for this work, we adopt the Tully–Fisher-based distance to NGC 2146, as it is tailored to the host galaxy and well-matched to the uncertainties inherent in SN II–based distances.

To estimate the total line-of-sight reddening toward SN~2024abfl, we measure the interstellar Na~ID absorption in the medium-resolution GMOS spectrum obtained on 2024 November 20. Both the Milky Way and host components are clearly detected. Using the Na~ID equivalent widths and applying Equation~(9) from \citet{2012MNRAS.426.1465P} together with the renormalization factor of 0.86 from \citet{2011ApJ...737..103S}, we derive a host-galaxy reddening of $E(B-V)_\mathrm{host} = 0.095 \pm 0.08$~mag and a Milky Way contribution of $E(B-V)_\mathrm{MW} = 0.186 \pm 0.08$~mag. These values combine to a total color excess of $E(B-V)_\mathrm{tot} = 0.28 \pm 0.11$~mag.

We note that the Milky Way value derived from Na~ID is higher than the dust-map estimate of \citet{2011ApJ...737..103S}. Given this discrepancy, and for the sake of consistency, we adopt the Na~ID--based reddening. Because Na~ID can saturate at moderate column densities, our Na~ID--derived values should be regarded as an upper limit on the true line-of-sight reddening.
Table~\ref{tab:properties} lists basic information and parameters of SN 2024abfl used throughout this work.


\begin{deluxetable}{lc}
\tablecaption{Properties of SN 2024abfl 
\label{tab:properties}}
\tablewidth{0pt}
\tablehead{
}
\startdata
RA (J2000) & 06$^{\mathrm{h}}$18$^{\mathrm{m}}$01\fs140\\
Dec. (J2000) & +78\degr22\arcmin01\farcs52\\
Last Nondetection (JD) & 2460626.76 \\
First Detection (JD) & 2460628.78 \\
Explosion Epoch (JD) & 2460628.41 \\
Redshift ($z$) & 0.002979 \\
Distance\textsuperscript{a} & $15.6^{+6.1}_{-3.0}$~Mpc \\
Distance modulus ($\mu$) & $30.97^{+0.71}_{-0.47}$~mag \\
$E(B-V)_\mathrm{MW}$\textsuperscript{b} & $0.19 \pm 0.08$~mag \\
$E(B-V)_\mathrm{host}$\textsuperscript{b} & $0.01 \pm 0.08$~mag \\
$E(B-V)_\mathrm{tot}$\textsuperscript{b} & $0.28 \pm 0.11$~mag \\
Peak Magnitude ($V_{\mathrm{max}}$) & $-14.87$ mag \\
Rise Time ($V$) & 4.32 days \\
\hline
\enddata
\tablenotetext{a}{From \citet{2021arXiv210912943C}.
\tablenotetext{b}{From the EW of Na ID lines.}}
\end{deluxetable}

\section{Observations and Data Reduction}
\label{sec:obs}
\subsection{Photometry}
After discovery, photometric observations were begun in the \textit{UBVgri} bands with the 0.4 and 1 m telescopes available through the Global Supernova Project under the Las Cumbres Observatory telescope network \citep{2013PASP..125.1031B}. PSF fitting was performed on the Las Cumbres data using \texttt{lcogtsnpipe}, a PyRAF-based photometric reduction pipeline \citep{2016MNRAS.459.3939V}. The $UBV$-band data were calibrated to Vega magnitudes \citep{2000PASP..112..925S} using standard fields observed on the same night by the same telescope. Finally, $gri$-band data were calibrated to AB magnitudes using the Sloan Digital Sky Survey \citep[SDSS,][]{2017ApJS..233...25A}.

UV and optical images were obtained at early epochs with the Ultraviolet/Optical telescope (UVOT; \citealp{2005SSRv..120...95R}) on board the Neils Gehrels Swift Observatory \citep[\textit{Swift}]{2004ApJ...611.1005G}. The data were downloaded from the NASA \textit{Swift} Data Archive\footnote{\url{https://heasarc.gsfc.nasa.gov/cgi-bin/W3Browse/swift.pl}}, and the images were reduced using standard software distributed with \texttt{HEAsoft}\footnote{\url{https://heasarc.gsfc.nasa.gov/docs/software/heasoft/}}. Photometry was performed for all the $uvw1$, $uvm2$, $uvw2$, $ U_\mathrm{S}$-, $ B_\mathrm{S}$-, and  $V_\mathrm{S}$-band images using a 3\farcs0 aperture at the location of SN 2024abfl. 
The contribution from the host galaxy, which was insignificant at that location, has not been subtracted. 

\begin{figure*}
    \centering
    \includegraphics[width=\linewidth]{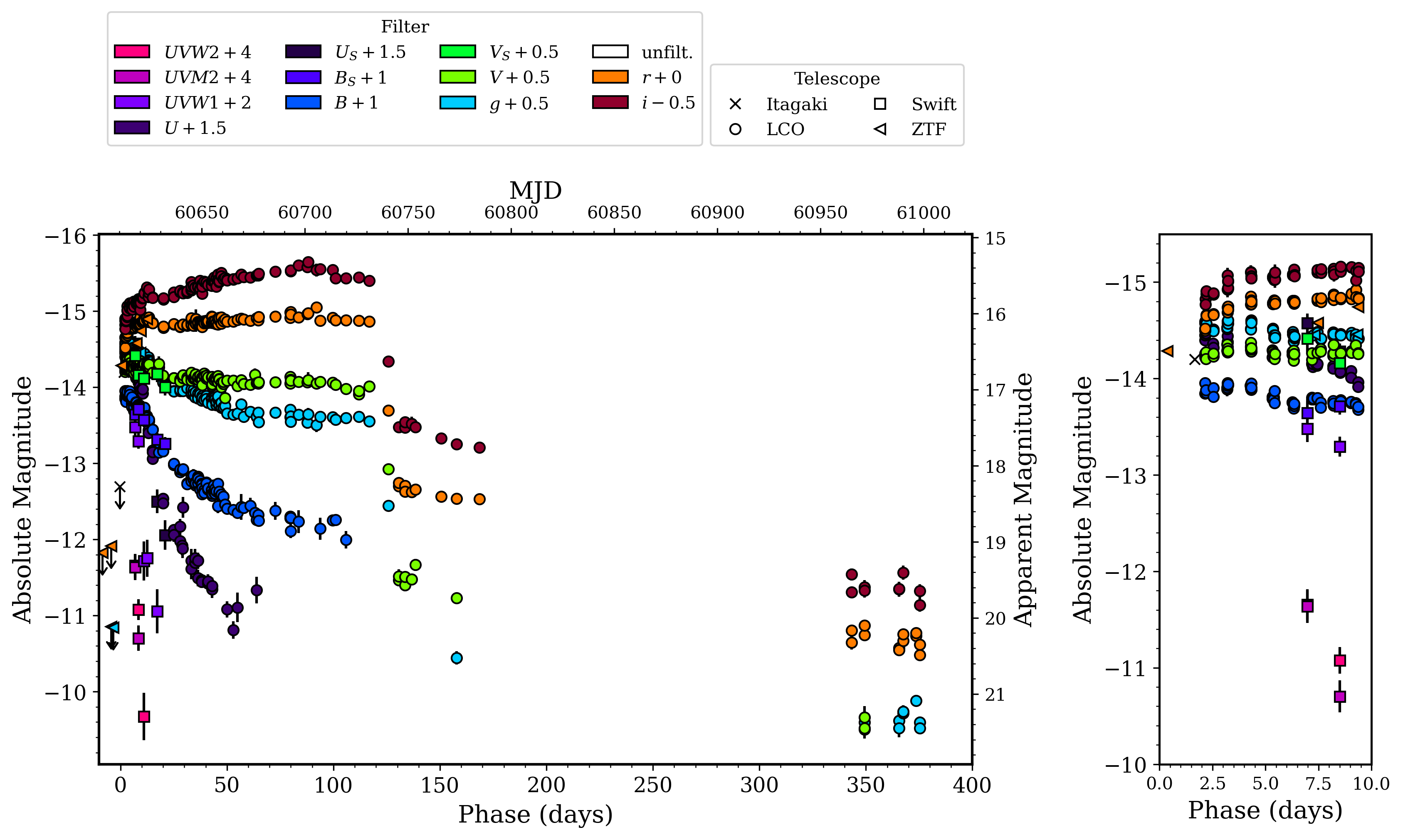}
    \caption{Optical photometry of SN 2024abfl with offsets and telescope indicated in the legend. The adopted explosion epoch is MJD 60627.91. The right panel expands the first week.}
    \label{fig:lightcurve}
\end{figure*}

In addition, we obtained photometry from the Zwicky Transient Facility \citep[ZTF;][]{2019PASP..131a8002B}, including both detections and deep non-detections. These measurements provide valuable constraints on the explosion epoch, enabling a more precise determination of the time of first light than would be possible using the discovery photometry alone. All of the resulting light curves are shown in Figure \ref{fig:lightcurve}.

\subsection{Spectroscopy}
Spectroscopic observations of SN 2024abfl began on 2024 November 16, approximately two days after discovery. Multiple epochs of optical spectroscopy were obtained at the Gemini North Observatory using the Gemini Multi-Object Spectrograph \citep[GMOS;][]{Hook2004} and reduced with the Data Reduction for Astronomy from Gemini Observatory North and South \citep[\texttt{DRAGONS};][]{Labrie2019} package following standard long-slit reduction procedures.

Additional spectra were acquired with the FLOYDS spectrograph \citep{2013PASP..125.1031B} on the 2-m Faulkes Telescope North as part of the Global Supernova Project. The data were reduced using the automated FLOYDS pipeline \citep{2014MNRAS.438L.101V} for order rectification, wavelength calibration, and flux calibration. 

Several epochs were obtained with the Multi-Object Double Spectrographs \citep[MODS;][]{Pogge2010} on the Large Binocular Telescope (LBT) located on Mt. Graham, Arizona USA. These observations are bias and flat-field corrected using the modsCCDred package \citep{Pogge2019}, then extracted and flux calibrated with standard IRAF routines.

Observations taken with Binospec \citep{Fabricant_2019} on MMT are initally processed using the Binospec IDL pipeline \citep{Kansky2019} for flat-fielding, sky subtraction, and wavelength and flux calibrations. The 1D spectrum is then extracted using IRAF techniques \citep{Tody1986, Tody1993}.

The spectrum taken with the Boller and Chivens Spectrograph (B\&C) on the University of Arizona’s Bok 2.3m telescope located at Kitt Peak Observatory is reduced using standard IRAF routines \citep{Tody1986, Tody1993}.
A log of the optical spectroscopic observations is provided in Table~\ref{tab:optspec}.

Two epochs of near-infrared spectroscopy were obtained with GNIRS on Gemini-N \citep{2006SPIE.6269E..4CE,2006SPIE.6269E..14E} in the cross-dispersed mode using the 32 line mm$^{-1}$ grating and a 0.65 arcsec-wide slit.  Additional NIR observations were acquired with the MMT and Magellan Infrared Spectrograph \citep[MMIRS;][]{2012PASP..124.1318M}. Both sets of data used a nearby A0V star for telluric correction and were taken in the standard ABBA pattern. GNIRS data were reduced using the IRAF GNIRS package, while MMIRS data were reduced using the MMIRS pipeline with additional manual processing, following procedures similar to those used for the MMT optical data. A summary of the NIR observations is given in Table~\ref{tab:nirspec}.

\section{Photometric Evolution}
\label{sec:phot}
\subsection{Optical and UV Lightcurves}
\begin{figure}
    \centering
    \includegraphics[width=\columnwidth]{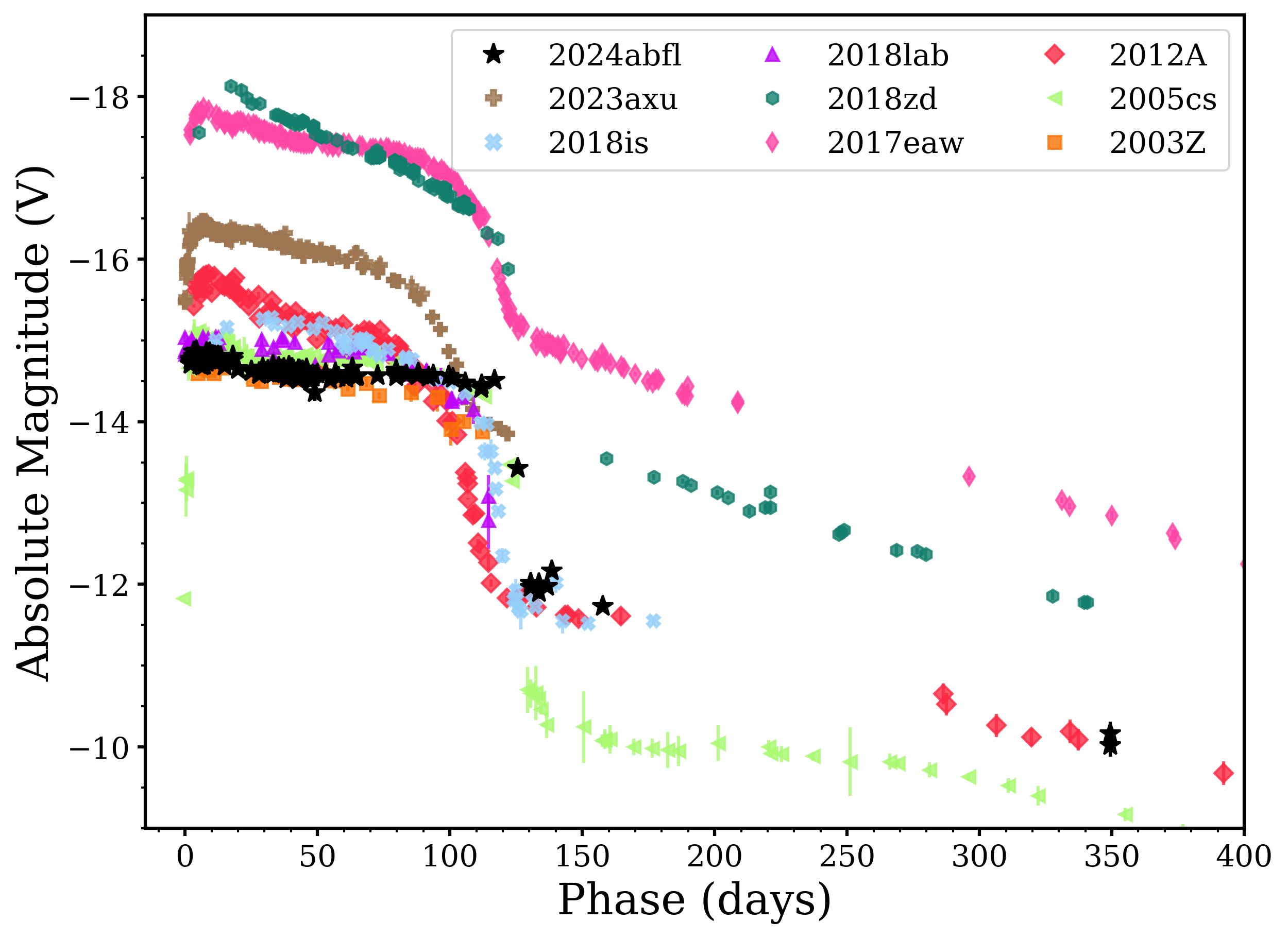}
    \caption{Absolute \textit{V}-band light curve of SN 2024abfl compared with other Type II SNe: 2003Z \citep{2014MNRAS.439.2873S}, 2005cs \citep{2009MNRAS.394.2266P}, 2012A \citep{2013MNRAS.434.1636T}, 2017eaw \citep{2019ApJ...876...19S}, 2018is \citep{2025A&A...694A.260D}, 2018lab \citep{2023ApJ...945..107P}, 2018zd \citep{2020MNRAS.498...84Z, 2022AstBu..77..407T}, and 2023axu \citep{2024ApJ...961..247S}. SN 2024abfl has a peak \textit{V}-band magnitude of $-14.87$ mag.}
    \label{fig:absV}
\end{figure}

We obtained photometric coverage from approximately 2 to 170 days post-explosion, before the SN became Sun-constrained, and then again starting around 325 days. The full optical and UV light curves of SN 2024abfl are shown in Figure \ref{fig:lightcurve}, and the absolute \textit{V}-band light curve, compared to other SNe IIP, is presented in Figure~\ref{fig:absV}. The \textit{V}-band light curve reaches a peak absolute magnitude of $M_V = -14.87$ approximately 4 days after explosion. Following this peak, SN 2024abfl then enters a relatively flat, extended plateau phase, as commonly seen in LLSNe, before declining into a faint radioactive tail phase. Plateau durations longer than $\sim$100 days are characteristic of LLSNe, where slower expansion velocities lead to higher ejecta densities and slower inward propagation of the recombination wave \citep{2014ApJ...786...67A, 2014MNRAS.442..844F, 2016AJ....151...33G}. 

As shown in Figure~\ref{fig:lightcurve}, the plateau of SN 2024abfl is notably flat in $V$ and $r$. This is even more noticeable in Figure~\ref{fig:absV} when compared to many other Type IIP SNe. The plateau duration and luminosity of SN~2024abfl is similar to that of SN~2005cs, SN~2003Z, and SN~2018lab corroborating its classification as a low-luminosity event. Although the transition from the plateau to the nebular phase is not densely sampled in our \textit{V}-band observations, the tail luminosity appears consistent with those of SNe 2012A and 2018is, and when we observe  SN~2024abfl again around day 350 the V band mag is consistent with SN~2012A. Figure~\ref{fig:absV} further illustrates that SNe II with comparable plateau luminosities can exhibit a wide range of tail luminosities, reflecting differences in the amount of $^{56}$Ni synthesized during the explosion.

Using Equation (1) from \citet{2016MNRAS.459.3939V}, we fit the \textit{V}-band light curve to derive parameters for comparison with other LLSNe. From the best-fit model, we determine $t_{\mathrm{PT}}$, the time from explosion to the midpoint between the end of the plateau and the onset of the radioactive tail, to be $125.99~\pm~0.25$ days. The parameter $w_0$ (in days) quantifies the steepness of the plateau-to-tail transition. Following \citet{2025A&A...694A.260D}, multiplying $w_0$ by six, yields a drop duration of $\sim$5.5 days. We find $a_0 = 2.412~\pm~0.031$ mag, which measures the depth of the plateau–tail drop. This value is typical for normal SNe IIP \citep{2010ApJ...715..833O}, although LLSNe can show deeper drops of $\sim$3–5 mag \citep{2014MNRAS.439.2873S, 2016MNRAS.459.3939V}, consistent with the diversity illustrated in Figure~\ref{fig:absV}. 
The \textit{V}-band decline rate of SN 2024abfl in the 50 days following maximum brightness, denoted $s_{50}$, is measured to be 0.274 mag/50d, slower than typical SNe IIP. 

\begin{figure}
    \centering
    \includegraphics[width=\columnwidth]{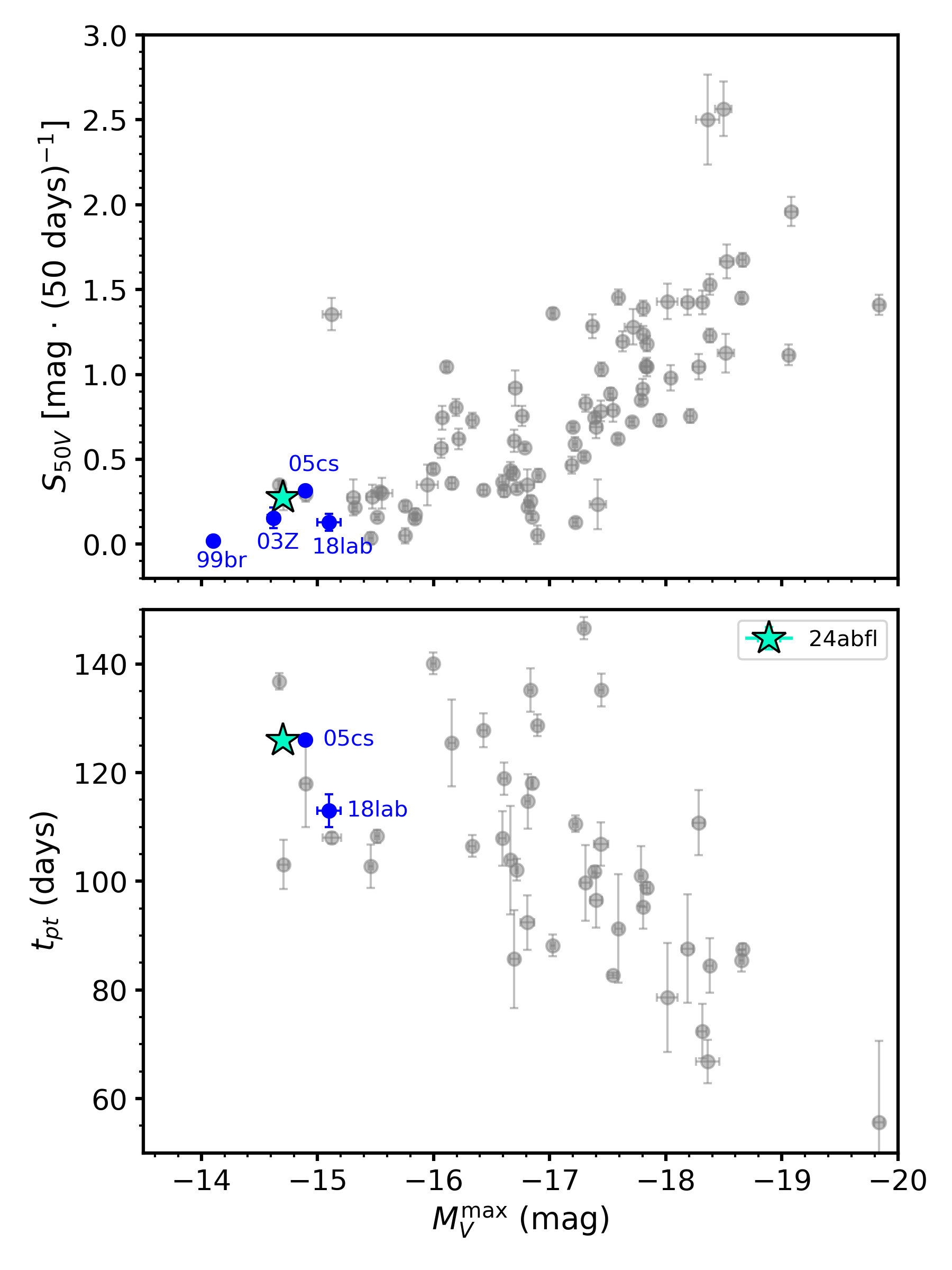}
    \caption{SN 2024abfl shown in the context of the SN II population from \citet{2016MNRAS.459.3939V}, plotted as peak absolute \textit{V-}band magnitude versus plateau decline rate ($S_{50V}$; top) and plateau duration ($t_{\mathrm{PT}}$; bottom). Several well-studied LLSNe are highlighted in blue for reference.”}
    \label{fig:valenticomp}
\end{figure}

In Figure~\ref{fig:valenticomp}, we compare SN 2024abfl to the SNe II sample from \citet{2016MNRAS.459.3939V} in terms of peak absolute \textit{V}-band magnitude against $s_{50V}$ and $t_{{PT}}$. SN 2024abfl occupies a region of parameter space populated by other well-studied LLSNe, including 1999br, 2003Z, 2005cs, and 2018lab, all of which exhibit faint peak magnitudes, extended plateaus, and shallow decline rates. The placement of SN 2024abfl reinforces the observed trend that LLSNe tend to show flatter and longer plateaus \citep{2014ApJ...786...67A, 2016MNRAS.459.3939V}.

\subsection{Color Evolution}
\begin{figure}
    \centering
    \includegraphics[width=\columnwidth]{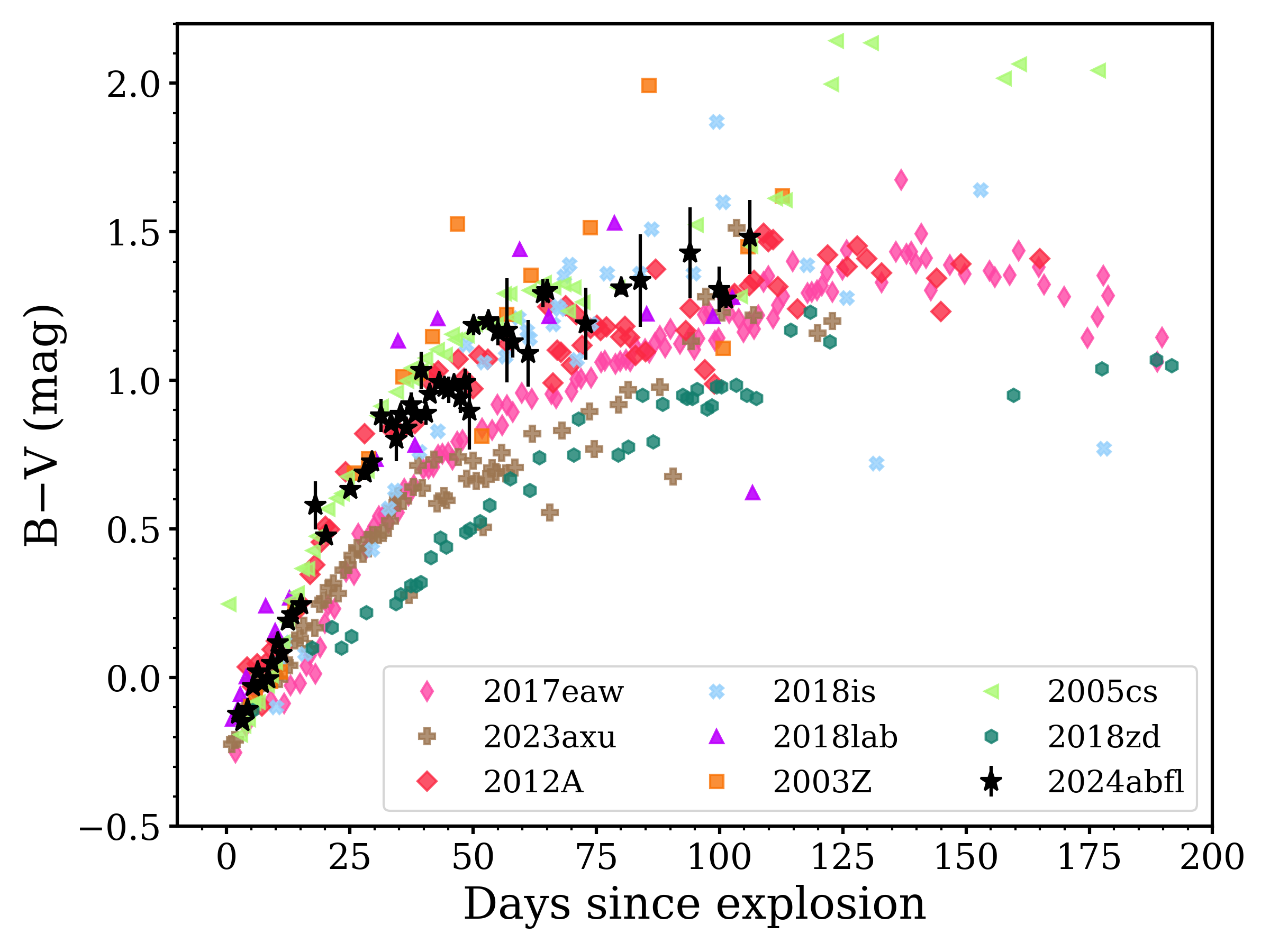}
    \caption{Extinction corrected $B-V$ color for SN 2024abfl compared with other SNe II, including other LLSNe. The adopted $E(B-V)_\mathrm{tot} =0.28$ mag is consistent with the color evolution of these similar SNe.}
    \label{fig:bv}
\end{figure}

In Figure \ref{fig:bv}, we show the $B-V$ color evolution of SN~2024abfl along with other normal and low luminosity Type II SNe. Following a continual reddening during the first $\sim$50 days, the $B-V$ color remains approximately constant through day $\sim$120, coinciding with the end of the plateau phase. After the recombination phase, SN~2024abfl appears to shift toward bluer colors. Overall, SN~2024abfl exhibits colors that lie on the red end of the sample distribution, consistent with the behavior of LLSNe  when compared to normal Type II SNe \citep{2009MNRAS.394.2266P, 2014MNRAS.439.2873S}.

\subsection{Bolometric Lightcurve and $^{56}$Ni Mass} 
\begin{figure}
    \centering
    \includegraphics[width=\columnwidth]{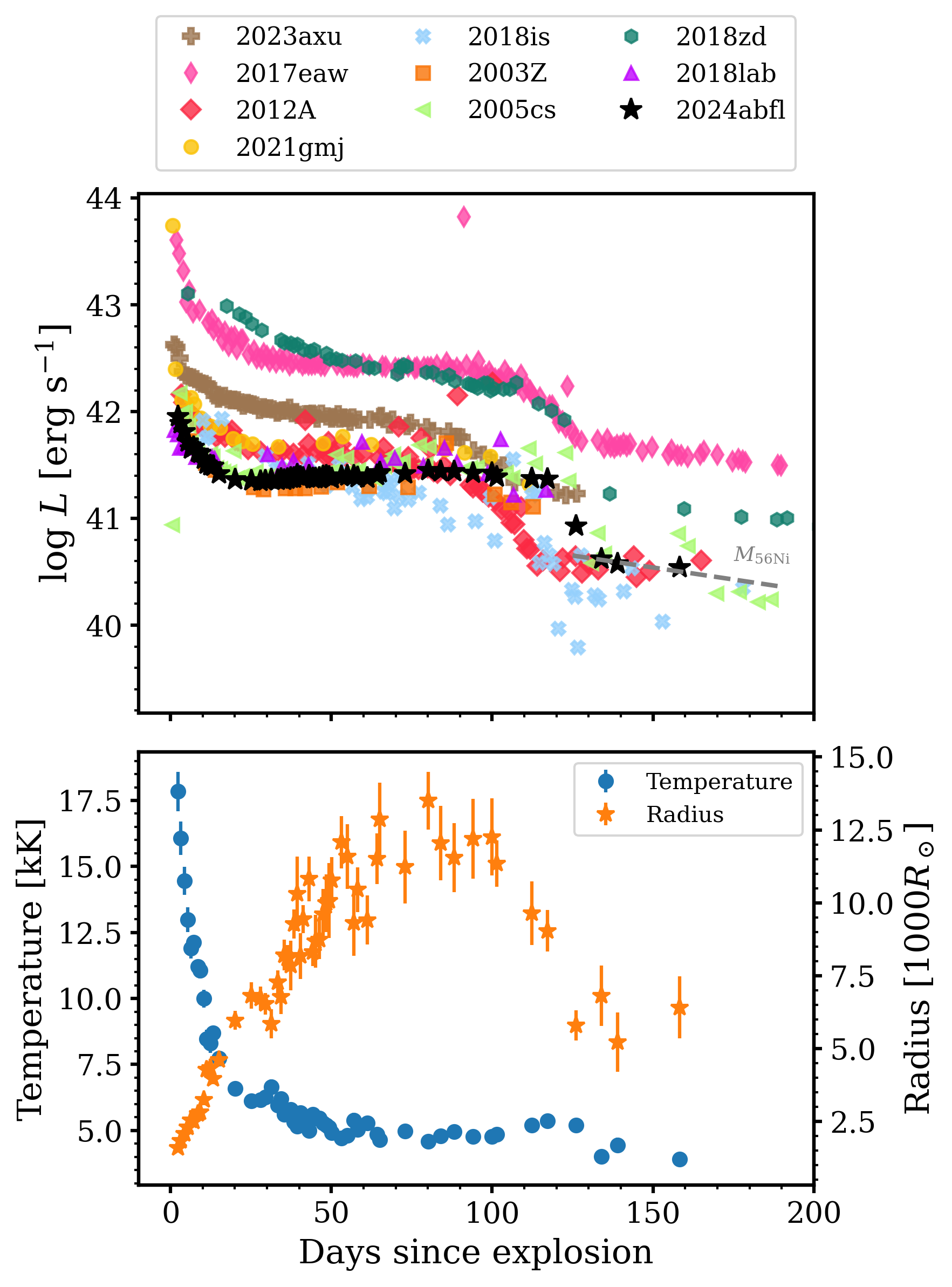}
    \caption{Top: Bolometric lightcurve of SN 2024abfl integrated from near-UV to NIR compared with other well-studied SNe II. The $^{56}$Ni decay rate is indicated in the gray dashed line. Bottom: Temperature and radius evolution of SN 2024abfl obtained from fitting the Planck function to the photometry using an MCMC routine. All data have been dereddened by our assumed $E(B-V)_\mathrm{tot} = 0.28$.}
    \label{fig:bolometric}
\end{figure}

The bolometric lightcurve of SN 2024abfl, calculated using the Light Curve Fitting package  \citep{2023zndo...7872772H}, is shown in the top panel of Figure \ref{fig:bolometric}. To construct the bolometric lightcurve, a Markov Chain Monte Carlo (MCMC) fitting routine fits a blackbody spectrum to each epoch of the observed SED. Assuming the reddening and distance listed in Table \ref{tab:properties}, we estimate a maximum bolometric luminosity of $L_{bol} = 9.02 \times 10^{41}$ erg s$^{-1}$, which is low even compared to other LLSNe, highlighting the particularly faint nature of this event.

The late-time bolometric luminosity appears to decline at a rate of 0.0075 mag day$^{-1}$, slightly slower than the canonical decay rate of fully trapped $^{56}$Co, which is 0.0098 mag day$^{-1}$. This is shown as the dashed line in the bottom of Figure \ref{fig:bolometric}, where we show the decay of a mass of 0.010~$M_\odot$ $^{56}$Ni for comparison. 

For our $^{56}$Ni mass estimates, we use the calculations described in \citet{2003ApJ...582..905H} and \citet{2012A&A...546A..28J}, both of which employ the radioactive tail of the bolometric light curve. We estimate the $^{56}$Ni mass at multiple epochs and find a value of $\sim0.010 \pm 0.001~M_\odot$. This is consistent with other LLSNe~II, although it is not as low as in very low-energy explosions, such as SN~2005cs, which produced only $\sim$0.002~$M_\odot$ of $^{56}$Ni \citep{2016MNRAS.459.3939V}. Our estimate is also in good agreement with the recent radiation-hydrodynamical modeling of \citet{2025arXiv250620068D}, who find $M_{\rm Ni} = 1.966^{+1.494}_{-1.736}\times10^{-2}~M_\odot$ for SN~2024abfl, consistent with our value within the combined uncertainties.

We also calculate the temperature and photospheric radius using the Light Curve Fitting package, shown in the bottom panel of Figure \ref{fig:bolometric}. As the SN expands and cools, the photospheric temperature and radius evolve over time. SN 2024abfl shows a gradual temperature, $T_{BB}$, decline over the first 30 days, from a peak temperature of $17.84 \pm 0.74$ kK at $\sim$2 days post explosion. This is followed by a slower decline from around 6000 to 4000 K over the next 100 days before the radioactive tail phase begins and the temperatures drop to 3000 K. This temperature evolution is consistent with that observed in other well-studied LLSNe, including SN~2005cs \citep{2014MNRAS.439.2873S} and SN~2018is \citep{2025A&A...694A.260D}, which show a similar two-phase decline characterized by an initial rapid cooling followed by a more gradual evolution during the plateau phase.

The photospheric radius, $R_{BB}$, increases to a maximum of approximately $13.5 \pm 0.99 \times 10^3~R_\odot$ at 80 days after explosion. This behavior is comparable to that seen in SN~2005cs, SN~2018is, and SN~2016bkv, which exhibit rapid early expansion followed by slower radius growth during the plateau. Overall, the temperature and radius evolution of SN~2024abfl fall well within the range observed for the LLSNe population.

\subsection{Early Lightcurve Modeling}
\begin{figure*}
    \centering
    \includegraphics[width=0.9\linewidth]{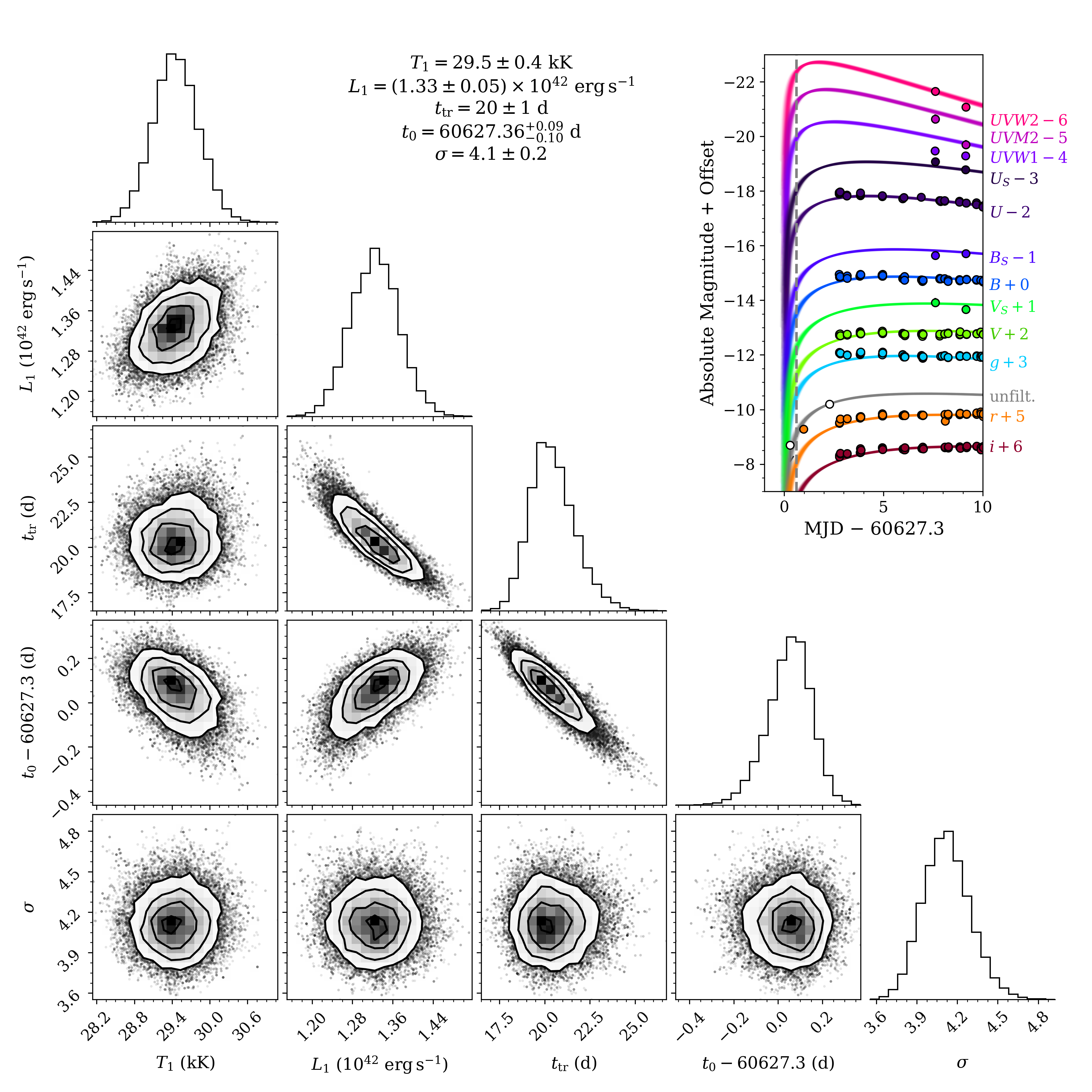}
    \caption{Posterior distributions of and correlations between the temperature and luminosity one day after explosion ($T_1$, $L_1$), the time at which the envelope becomes transparent ($t_{\mathrm{tr}}$), and the explosion time ($t_0$). The top-right panel shows a sample of model light curves using randomly selected parameters drawn from the posterior, together with the observed magnitudes. The resulting median values and 1$\sigma$ uncertainties for the best-fit parameters are listed at the top of the figure. The model does not fit well with the UV bands and it violates our explosion epoch (vertical gray line) constrained by nondetections, indicating that it is unable to fit the fast rise in the light curves.}
    \label{fig:shockcooling}
\end{figure*}

The early photometric evolution of Type II SNe is thought to be partially powered by shock cooling, during which energy added to the stellar envelope by the outgoing shock is released as the material expands and thins. Because SN 2024abfl was observed with exceptionally high cadence during the first several days after explosion, we model its early-time light curves using the Light Curve Fitting package \citep{2023zndo...7872772H}, which implements the analytic shock-cooling framework of \citet{2017ApJ...838..130S}.

We adopt the version of the Sapir $\&$ Waxman prescription employed by \citet{2018ApJ...861...63H}, which is parameterized by the temperature one day after explosion ($T_1$), the bolometric luminosity one day after explosion ($L_1$), the time at which the envelope becomes transparent ($t_{\mathrm{tr}}$), and the explosion time ($t_0$). This formulation assumes a polytropic index of $n = 1.5$, corresponding to an RSG progenitor envelope structure.
Sapir $\&$ Waxman's models are valid starting when the supernova shock traverses the progenitor radius ($t > \frac{R}{5v}$) and ending when the ejecta cool below 8000 K. To remain within these limits, we restrict the fit to epochs earlier than MJD 60640.0 (11.8 days after explosion). 

The high-cadence optical coverage for SN~2024abfl provides tight constraints on the shock-cooling slope. We fit the multi-band light curve using an MCMC sampler with flat priors on all parameters, bounded to ensure physically allowed temperatures, luminosities, and explosion times consistent with our earliest nondetections. The model gives the bolometric luminosity and blackbody temperature as a function of time, which are then converted to observed fluxes at each photometric epoch. Figure~\ref{fig:shockcooling} shows the results of the MCMC analysis, including the early light-curve fits, posterior parameter distributions, and 1$\sigma$ credible intervals centered on the median values.

One day after explosion, the modeled temperature is $(29.5 \pm 0.4)\times10^3$ K (kK) with a luminosity of $(1.33 \pm 0.05) \times10^{42}$ erg s$^{-1}$. The best-fit explosion time is MJD $60627.3 \pm 0.1$ d. This is less than one day earlier than our adopted explosion epoch (MJD 60627.91) and slightly earlier than the explosion window allowed by the nondetections and first detection (MJD 60627.59–60628.78).

The inability of the shock-cooling model to reproduce the steep early rise may indicate the presence of ejecta–CSM interaction, which is not accounted for in the Sapir $\&$ Waxman framework. Similar discrepancies between analytic shock-cooling models and early light curves have been reported for other SNe with early-time observations \citep[e.g.,][]{2023ApJ...953L..16H, 2023ApJ...945..107P, 2024ApJ...972L..15S, 2024ApJ...965...85A, 2024ApJ...970...96I}. Even relatively low-density CSM can contribute additional luminosity through early ejecta–CSM interaction, producing a faster rise than expected from a bare RSG envelope. In this scenario, interaction with extended material above the stellar surface provides an additional source of radiative energy as the ejecta decelerate and cool, leading to an enhanced early-time luminosity \citep{2017ApJ...838...28M, 2022MNRAS.512.2777T}. This interpretation is further supported by the presence of broad early-time emission features detected in spectra obtained within three days of explosion (Section~\ref{sec:flash}), consistent with interaction with low-density, rapidly overrun CSM rather than long-lived, narrow-line interaction.

\subsection{1D Hydrodynamical 
Modeling}
\label{sec:hydro}
We use the open-source 1D radiation hydrodynamics code, Supernova Explosion Code \citep[\texttt{SNEC};][]{2015ApJ...814...63M}, for light curve modeling to infer the progenitor parameters and explosion properties of SN 2024abfl, as was done for SN~2018is \citep{2025A&A...694A.260D}. \texttt{SNEC} uses the inputs of $^{56}$Ni mass, and $^{56}$Ni mixing, explosion energy, and progenitor models and produces multi-band and bolometric light curves, photospheric velocity, and temperature evolution. Similarly to \citet{2025A&A...694A.260D} we use the RSG models from \citet{2016ApJ...821...38S} for the progenitor. The distance and reddening are fixed with the values listed in Table 1 and with a $^{56}$Ni mass of 0.01 M$_{\odot}$. 

\begin{figure}
    \centering
    \includegraphics[width=\columnwidth]{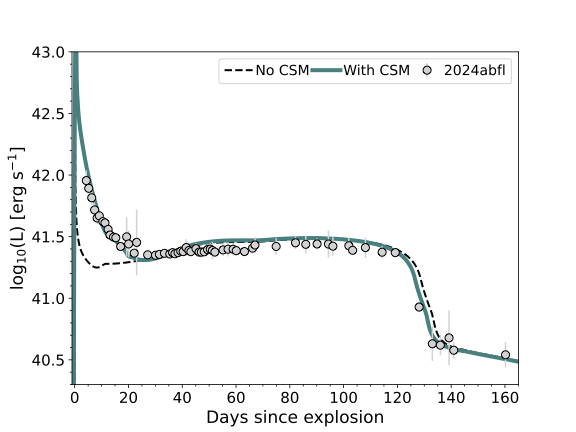}
    \caption{Model bolometric luminosities for scenarios with and without CSM.}
    \label{fig:SNEClc}
\end{figure}

\begin{figure}
    \centering
    \includegraphics[width=\columnwidth]{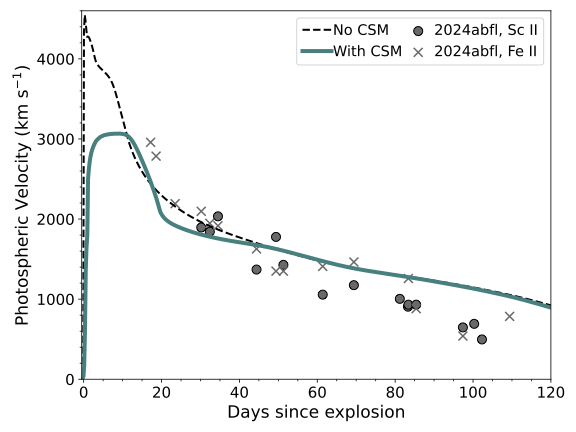}
    \caption{Model photospheric velocities for scenarios with and without CSM compared to observed line velocities of \ion{Sc}{2} $\lambda6246$ and \ion{Fe}{2} $\lambda5169$.}
    \label{fig:SNECvel}
\end{figure}

We show in Figures \ref{fig:SNEClc} and \ref{fig:SNECvel} the bolometric luminosities and photospheric velocities from \texttt{SNEC} compared with observational data of SN~2024abfl. For both fits, models were produced with no CSM and with 0.1 M$_{\odot}$ of CSM, attached to the progenitor surface and extending up to 1200 R$_\odot$. The model with additional CSM does a better job of fitting the bolometric luminosity at early times. The length of the plateau is suitably reproduced, suggesting that the ZAMS mass of the progenitor and explosion energy are fairly well constrained. 

The best fit corresponds to a 9.0 $M_{\odot}$ ZAMS star, with a pre-SN mass of 8.75 M$_{\odot}$, a pre-SN stellar radius of 403 R$_{\odot}$, and an explosion energy of 0.12 foe. For comparison, \citet{2025arXiv250620068D} used radiation-hydrodynamical models from \citet{2023PASJ...75..634M} to model the progenitor properties of LLSNe and found that the best match for the progenitor of SN~2024abfl is a $10.75\pm{0.45}$ M$_{\odot}$ ZAMS star with an explosion energy between 0.10-0.14 foe, which is consistent with our values. Both progenitor mass estimates agree with the value found using the archival HST data \citep{2025ApJ...982L..55L}.

\section{Spectroscopic Evolution}
\label{sec:spec}
\subsection{Overall Evolution}
\label{sec:spec1}
\begin{figure*}
    \centering
    \includegraphics[width=\linewidth]{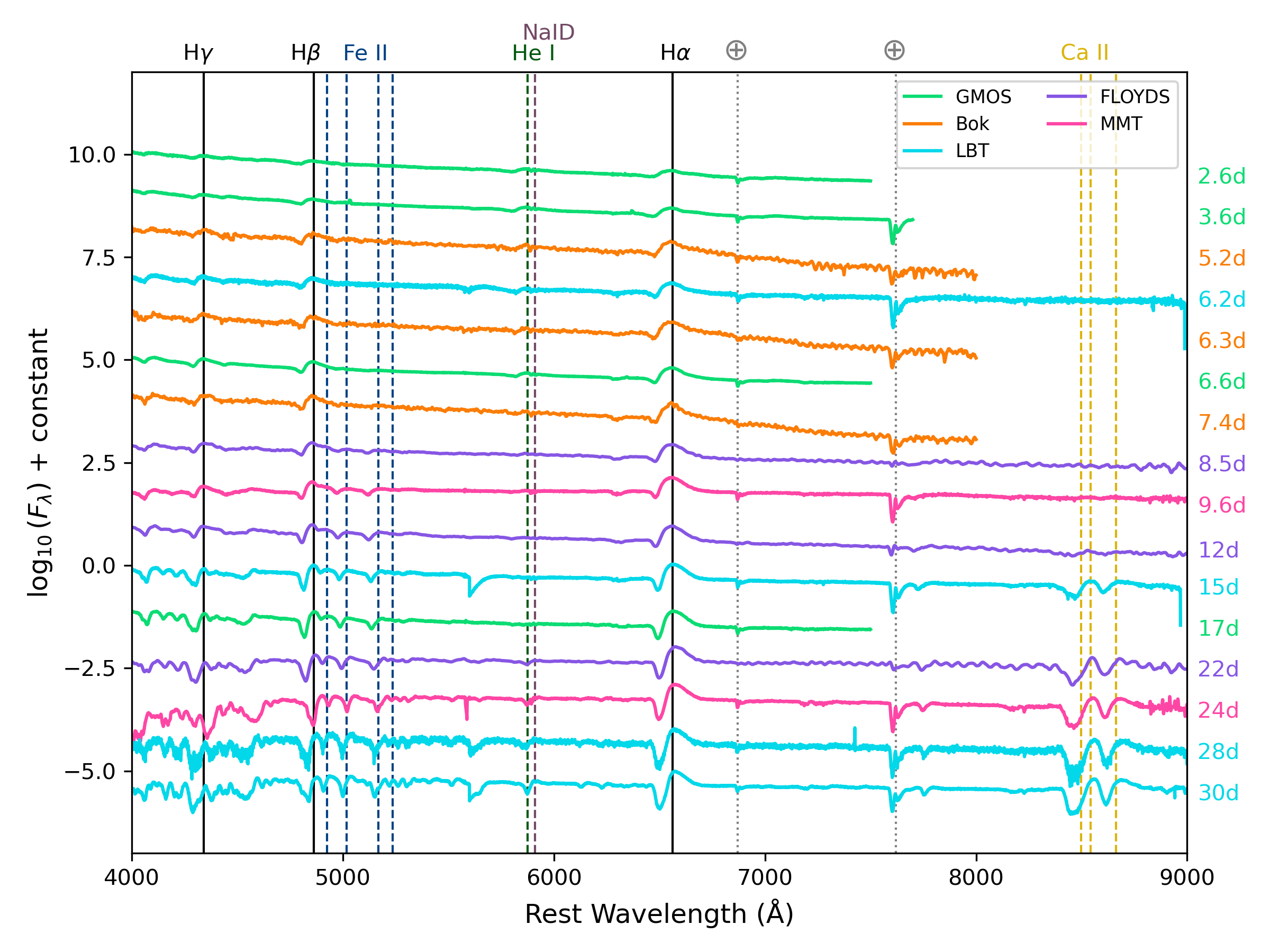}
    \caption{Optical spectral sequence of SN 2024abfl for the first 30 days post explosion, shifted vertically for clarity. Each telescope+instrument pair is denoted by a different color. The dates are with respect to our assumed explosion epoch of MJD 60627.91.}
    \label{fig:earlyspec}
\end{figure*}

\begin{figure*}
    \centering
    \includegraphics[width=\linewidth]{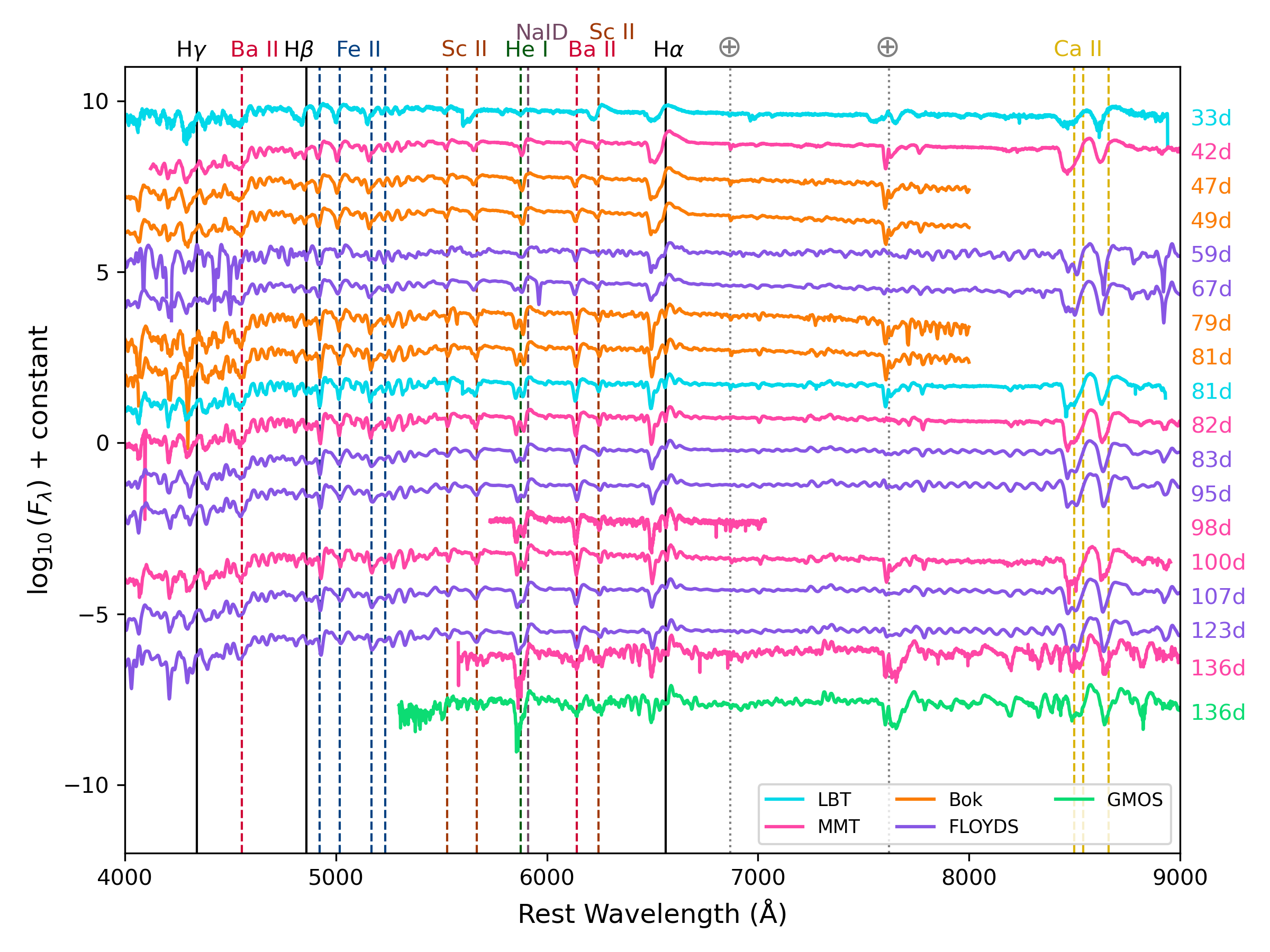}
    \caption{The same as Figure \ref{fig:earlyspec}, but for 33\text{--}136 days after explosion.}
    \label{fig:latespec}
\end{figure*} 

The optical spectra of SN 2024abfl, obtained during the first 140 days after explosion, are presented in Figures \ref{fig:earlyspec} and \ref{fig:latespec}. The spectral evolution of SN 2024abfl is similar to that of other LLSNe presented in previous papers \citep[eg.,][]{2001MNRAS.322..361B, 2004MNRAS.347...74P, 2009MNRAS.394.2266P, 2014MNRAS.439.2873S, 2022MNRAS.513.4983V}. In the first few days, the spectra exhibit a blue continuum with weak and shallow P-Cygni profiles of the Balmer H lines. These early time spectra show no signs of prominent, narrow emission flash lines often seen in SNe II shortly after explosion, although it is possible they could have disappeared before our first epoch of spectroscopy (at +2.6d). Instead, there is a broad emission feature around 4600 \AA, which will be further analyzed in Section \ref{sec:flash}.

Starting around day 15, \ion{Fe}{2} $\lambda4924$ and $\lambda5018$ lines emerge and can be used to trace the photospheric velocity, while at the same time the \ion{Ca}{2} infrared triplet ($\lambda8498$, $\lambda8542$, $\lambda8662$) becomes visible. Around day 24, more \ion{Fe}{2} lines ($\lambda5169$, $\lambda5234$) appear and the P-Cygni profiles of the hydrogen features become deeper and more prominent. Around a month after explosion, several metal lines become more prominent, including \ion{Sc}{2} $\lambda5527$, $\lambda5658$, $\lambda6246$ and \ion{Ba}{2} $\lambda6142$, which are commonly seen in LLSNe \citep{2004MNRAS.347...74P, 2014MNRAS.439.2873S, 2017ApJ...850...89G}. Between day 60 and day 80, prominent lines of \ion{Ba}{2} $\lambda6496$, \ion{Fe}{2}, and \ion{Sc}{2} develop blueward of H$\alpha$, and H$\alpha$ itself seems to show multi-component features. These lines and the H$\alpha$ evolution will be further analyzed in Section \ref{sec:halpha}.

\begin{figure*}
    \centering
    \includegraphics[width=\linewidth]{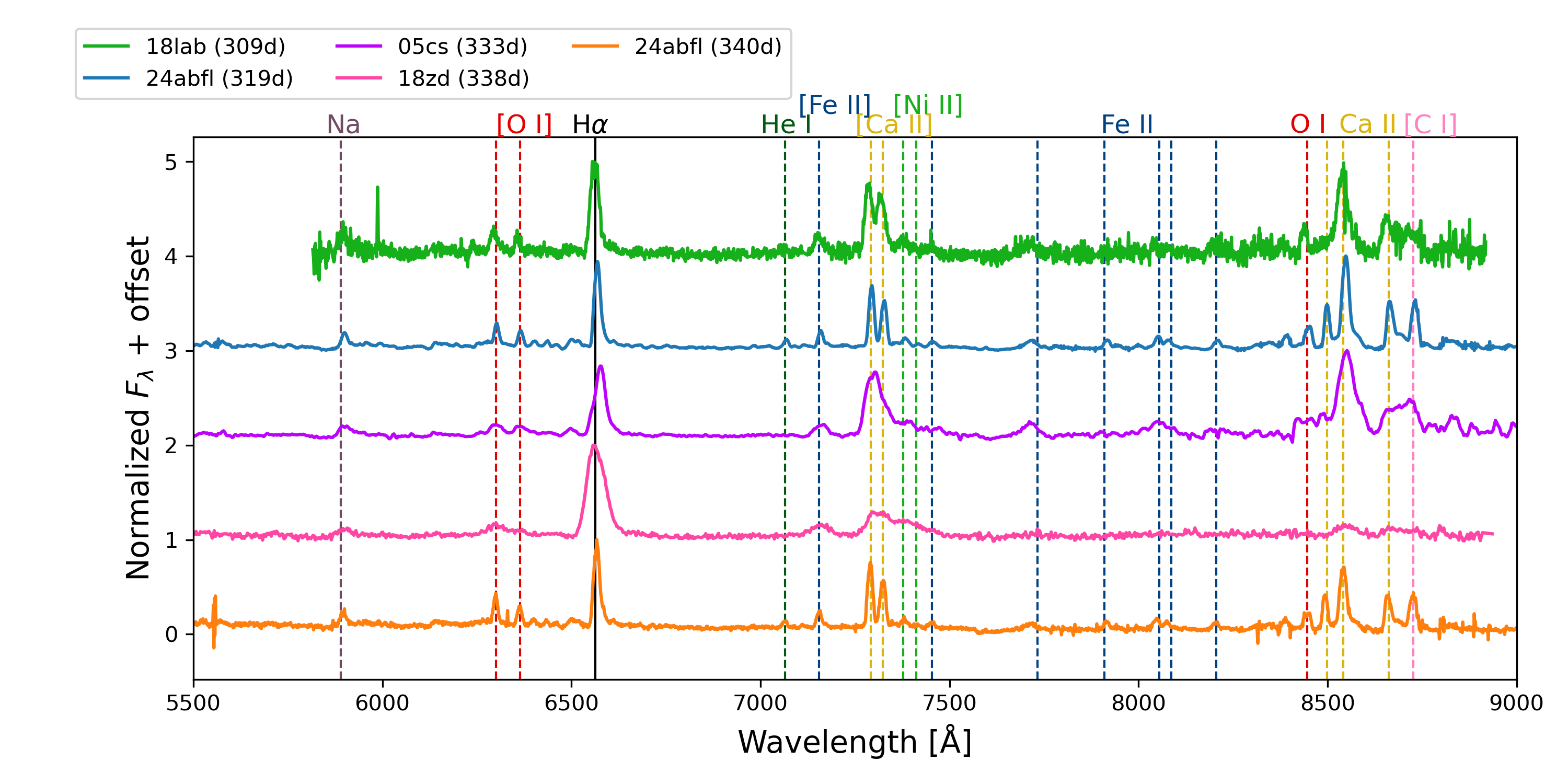}
    \caption{Late-time spectra of SN 2024abfl taken 319 and 340 days post explosion compared to the late time spectra of SN 2018lab \citep{2023ApJ...945..107P}, SN 2005cs \citep{2009MNRAS.394.2266P}, and SN 2018zd \citep{2021NatAs...5..903H}.}
    \label{fig:nebularcomp}
\end{figure*}

After the fall from the plateau, we obtained a spectrum at 136 days post explosion, which exhibits a transition from a P-Cygni-dominated spectrum to an emission-line nebular spectrum where we can observe lines such as \ion{O}{1}, \ion{He}{1}, and \ion{Fe}{2}. This marks the end of the photospheric phase of the SN, as the appearance of nebular emission lines indicates that the expanding ejecta have become optically thin. Our last two optical spectra were taken well into the nebular phase at 319 and 340 days post-explosion (some of the latest spectra of a LLSNe to be published) and are shown in Figure \ref{fig:nebularcomp}. The spectra are dominated by H$\alpha$, Ca, and O lines, with Na, Fe, and C as well.  The late-time spectra are discussed in detail in Section \ref{sec:Nebular}.

\subsection{Early `Ledge' Spectroscopy}
\label{sec:flash}
As discussed in Section \ref{sec:spec1}, SN 2024abfl does not exhibit the narrow, high-ionization emission lines that characterize the earliest spectra of interacting SNe II, even though we obtained a high signal to noise spectrum within 3 days of our inferred explosion epoch. Instead, the first two spectra ($<4$ days post-explosion) show a broad emission feature around 4600 \AA. This “ledge” feature peaks near \ion{He}{2} $\lambda4686$ in SN 2024abfl and is the most clearly visible in the spectrum 2.6 days post-explosion, though it is also present in the second spectrum (3.6 days post-explosion).

\begin{figure}
    \centering
    \includegraphics[width=\columnwidth]{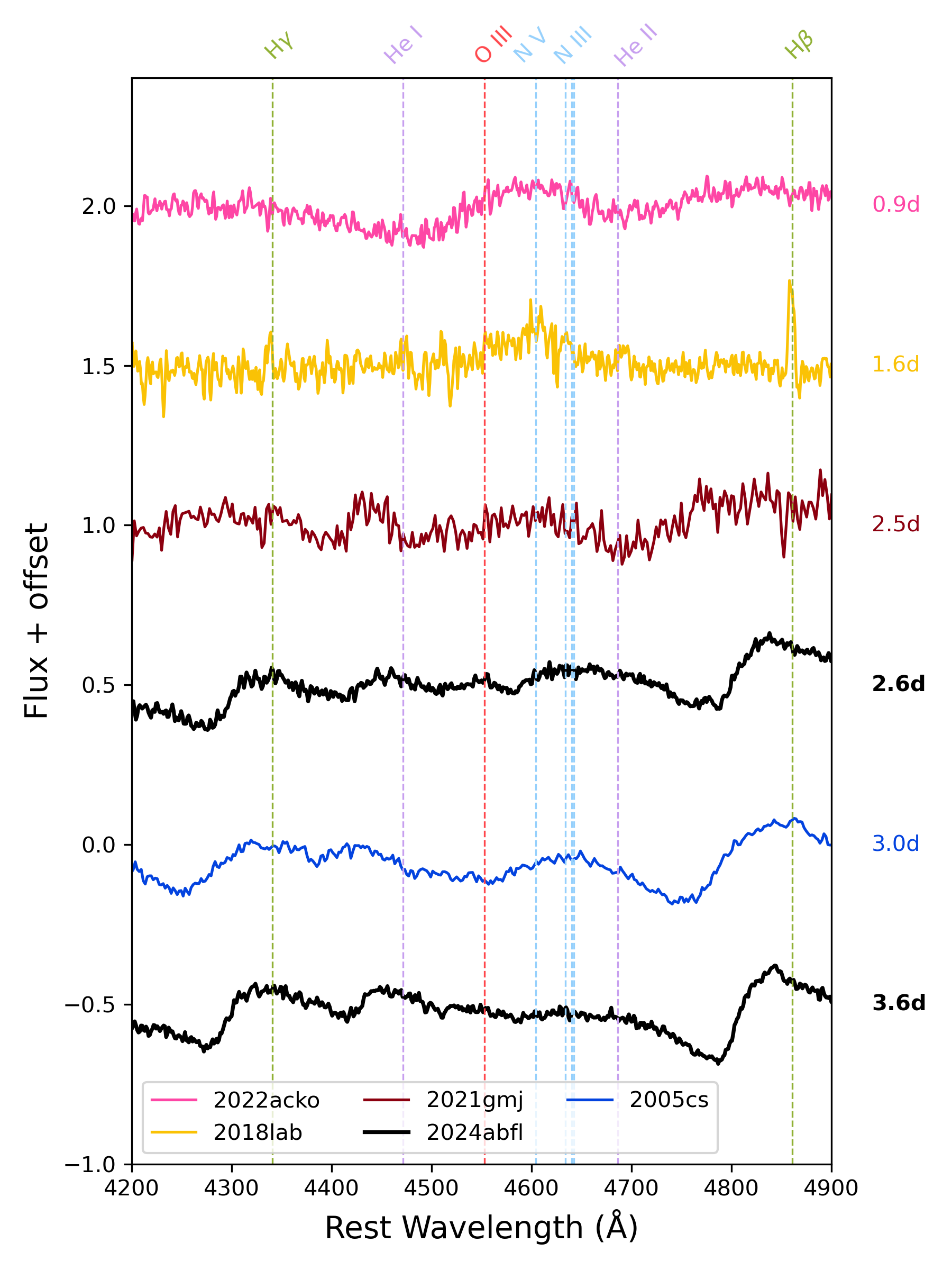}
    \caption{Comparison of the ledge feature at $\sim4600$ \AA\ seen in SN 2024abfl compared with other LLSNe with similar spectral profiles during the first $<$5 days after explosion: SN 2005cs \citep{2006MNRAS.370.1752P, 2009MNRAS.394.2266P}, SN 2016bkv \citep{2018ApJ...861...63H}, SN 2018lab \citep{2023ApJ...945..107P}, and SN 2022acko \citep{2025MNRAS.540.2591L}.}
    \label{fig:flash}
\end{figure}

Although this feature is sometimes grouped under the umbrella of “flash-ionization” signatures, we use the term “ledge” to distinguish this broad emission from the narrow, high-ionization lines that appear in very early spectra of interacting SNe II. Very few LLSNe have spectra $<5$ days following explosion. However, among those that do, such as SN 2005cs \citep{2009MNRAS.394.2266P}, SN 2016bkv \citep{2018ApJ...861...63H}, and SN 2018lab \citep{2023ApJ...945..107P}, the majority appear to exhibit a feature similar to what we observe for SN 2024abfl (see Figure \ref{fig:flash}).

Several interpretations have been proposed for this feature in past works. For SN 2005cs, which shows strong similarities to SN 2024abfl, \citet{2006MNRAS.370.1752P} discussed a possible identification as high-velocity (HV) H$\beta$, following earlier suggestions for SNe II. However, based on the inconsistent velocities of the supposed HV components and their simultaneous disappearance, they ultimately argued against a HV Balmer origin and instead favored an interpretation in terms of \ion{N}{2} (for the $\sim$4580 \AA\ feature) and \ion{Si}{2} $\lambda6355$ (for the $\sim$6300 \AA\ feature).

In our spectra of SN 2024abfl, we do observe a possible HV-like absorption blueward of H$\alpha$ at $\sim$3.6 days after explosion, but the overall evidence for genuine HV H$\beta$ is weak, consistent with the conclusions reached by \citet{2006MNRAS.370.1752P} for SN 2005cs.

A more compelling explanation is that the 4600 Å ledge arises from a blend of high-ionization shock accelerated CSM features, particularly \ion{N}{3} and \ion{He}{2} \citep{2017A&A...605A..83D}, an interpretation that has been invoked for several LLSNe including SN 2016bkv and SN 2018lab \citep{2020ApJ...902....6S, 2021ApJ...912...46B, 2022ApJ...935...31H, 2023ApJ...945..107P}. The relatively modest blueshift of the feature in SN 2024abfl compared to those events further supports a blended high-ionization origin rather than a HV Balmer component. Consistent with this picture, the possible absorption blueward of H$\alpha$ can plausibly be attributed to \ion{Si}{2} $\lambda6355$, as also favored by \citet{2006MNRAS.370.1752P} for SN 2005cs.

To further understand the origin of the ledge feature, we attempted to compare the first spectrum of SN 2024abfl with the model spectra of \citet{2017A&A...605A..83D}. In contrast to SN 2018lab and 2021gmj, which are reasonably well described by the extended-atmosphere models, SN 2024abfl shows a ledge that is not sufficiently blueshifted or asymmetric to conclusively match any of the current models.

The morphology of SN 2024abfl’s ledge feature highlights the significant diversity observed in the earliest spectra of SNe II. Symmetric narrow-lined flash features, like those seen in SN 2016bkv, are produced by non-coherent scattering of thermal electrons. In contrast, broad or asymmetric features may arise from bulk motions in material swept up by the expanding ejecta \citep{2009MNRAS.394...21D}. The broad, blended, ledge in SN 2024abfl lies firmly in this latter category, indicating the presence of some highly ionized and shock-accelerated CSM but at relatively low density or confined radial extent.

\subsection{H$\alpha$ Evolution}
\label{sec:halpha}
\begin{figure}
    \centering
    \includegraphics[width=\columnwidth]{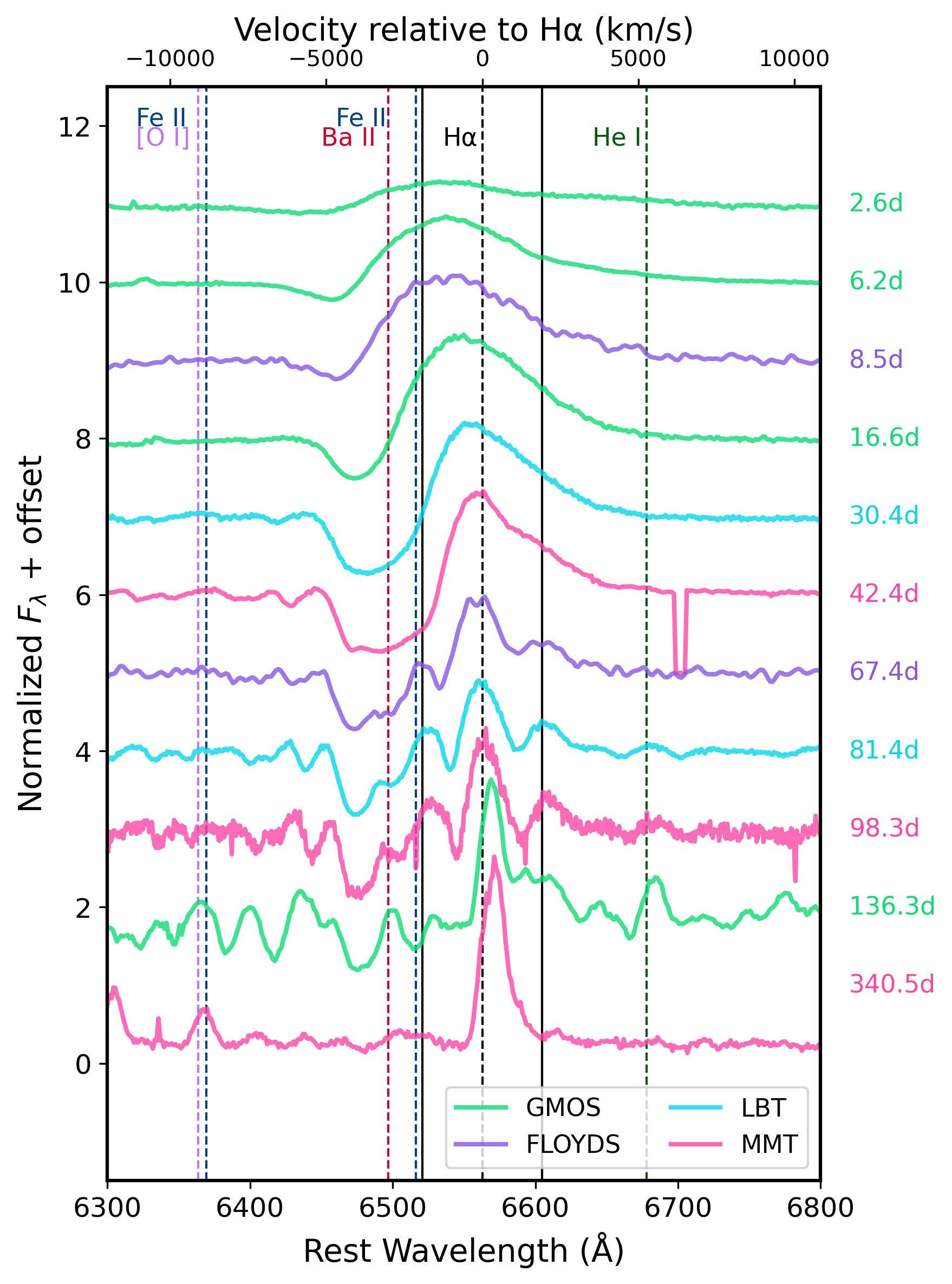}
    \caption{Evolution of H$\alpha$ and nearby lines starting at 2.6 days post explosion. Dashed vertical lines show the rest-frame wavelengths of key spectral features, while solid vertical lines mark symmetric velocity offsets of $\pm1900$ km s$^{-1}$ with respect to H$\alpha$.}
    \label{fig:halpaev}
\end{figure}

In Figure \ref{fig:halpaev}, we present the evolution of H$\alpha$ in SN 2024abfl from $\sim$2 to 340 d post-explosion, normalized to the local continuum at each epoch. From our first spectrum the start of a somewhat broad ($\sim$ 5000 km s$^{-1}$) and blueshifted H$\alpha$ emerges, and by a week post-explosion it begins to exhibit a distinct P-Cygni profile. The blueshifted peak migrates from $-$1500 km s$^{-1}$ to zero velocity at day 67 where it remains until the end of the plateau. This behavior is consistent with the expected outward cooling of the ejecta and the gradual exposure of deeper, slower-moving layers. \citep{2014MNRAS.441..671A}. 

The H$\alpha$ profile becomes complex starting in the day 59 spectrum, in the second half of the plateau phase (Figure \ref{fig:latespec}). This complex H$\alpha$ profile is not uncommon in LLSNe and has been described as the result of the combination of H$\alpha$ and \ion{Ba}{2} $\lambda$6497 \citep{2001MNRAS.322..361B, 2009MNRAS.394.2266P, 2014MNRAS.438..368T, 2017MNRAS.466...34L, 2022MNRAS.513.4983V}. However, in SN 2024abfl, the additional components are too strongly blueshifted and too symmetrical to be explained by \ion{Ba}{2} alone. Instead, they form a nearly symmetric pair of features at roughly $\pm1900$ km s$^{-1}$ around the rest wavelength of H$\alpha$. 

These features are prominent for about the last month of the plateau, but are substantially weaker by day 136 and appear to be non-existent by day 319. This could indicate interaction with a disc or torus of CSM surrounding the progenitor.  If we assume a shock velocity of 3000 km s$^{-1}$, by day 67 this would be interacting with material at roughly 2 $\times$ 10$^{15}$ cm away. Interaction with CSM in a disc or torus at late times has been seen in other normal luminosity Type IIP \citep[][]{, 2001ApJ...553..861L, 2010ApJ...715..541A, 2016MNRAS.457.3241A, 2025ApJ...980...37A, 2015MNRAS.449.1876S, 2023ApJ...956...46S} and flattened disk-like CSM is inferred to be common in SNe IIn based on polarization \citep{2024MNRAS.529.1104B}. Multi-peaked H$\alpha$ may have even been seen in other LLSNe but could have been thought to be lines of other elements \citep[Ti, for instance,][]{2004MNRAS.347...74P}. This may indicate that this phenomenon may be more common among LLSNe than previously appreciated, but remains insufficiently explored in the literature. 

\subsection{Line Velocity Evolution}
\begin{figure}
    \centering
    \includegraphics[width=\columnwidth]{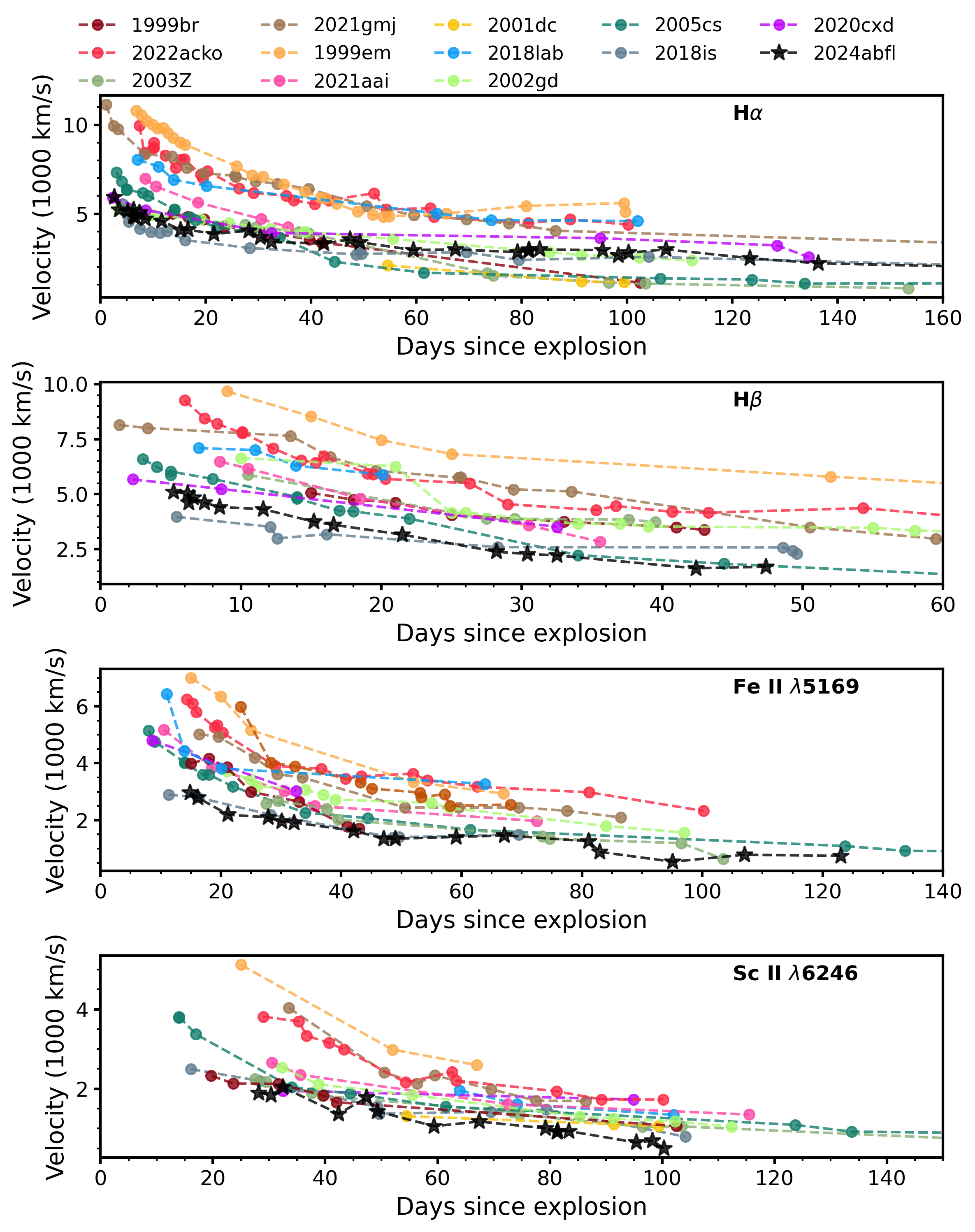}
    \caption{Velocity evolution of SN 2024abfl as a function of phase for H Balmer and metal lines, compared with a sample of low and standard luminosity SNe II-P adapted from \citet{2025A&A...694A.260D}.}
    \label{fig:velocity}
\end{figure}

The velocities of H$\alpha$, H$\beta$, \ion{Fe}{2} $\lambda5169$, and \ion{Sc}{2} $\lambda6246$ in the ejecta are estimated from the position of their absorption minima and shown against a comparison sample adapted from \citet{2025A&A...694A.260D} in Figure \ref{fig:velocity}. The velocity evolution of SN 2024abfl is remarkably similar to SN 2018is. Prior to 20 days, the H$\alpha$ velocity evolution of SN 2024abfl is among the lowest of the comparison sample before settling in around 3000 km s$^{-1}$ with little evolution after. The H$\beta$ feature evolves similarly, but settles at a lower velocity. SN 2002gd \citep{2014MNRAS.439.2873S}, 2018is \citep{2025A&A...694A.260D}, 2020cxd \citep{2022MNRAS.513.4983V}, 2021gmj \citep{2024ApJ...971..141M}, and 2022acko \citep{2023ApJ...953L..18B} also exhibit similar flattening in velocity after a more rapid initial decline. In the case of LLSNe, this flattening is likely due to the formation of \ion{Ba}{2} $\lambda6497$, which complicates the estimate of the true absorption minima of H$\alpha$ \citep{2025A&A...694A.260D}.

Compared to LLSNe, the expansion velocity of the metal lines \ion{Fe}{2} $\lambda5169$ and \ion{Sc}{2} $\lambda6246$ in SN 2024abfl is consistently lower across all epochs. The photospheric velocity, inferred from the \ion{Sc}{2} $\lambda6246$ minimum, decreases rapidly from $\sim$2000 km s$^{-1}$ at about a month after explosion to less than 700 km s$^{-1}$ at $\sim$100 days. While lower explosion energies can cause slower velocities, CSM interaction will also decelerate SN ejecta, such that high-density CSM may result in ejecta speeds $\sim$1000 km s$^{-1}$ slower than low-density CSM \citep{2017A&A...605A..83D}. However, because lower expansion speeds are characteristic of LLSNe, we are unable to set limits on the density of the CSM from the velocity measurement alone, but it does corroborate the possible asymmetric H$\alpha$ caused by CSM interaction.

\subsection{NIR Spectra}
\label{sec:NIR}
\begin{figure}
    \centering
    \includegraphics[width=\columnwidth]{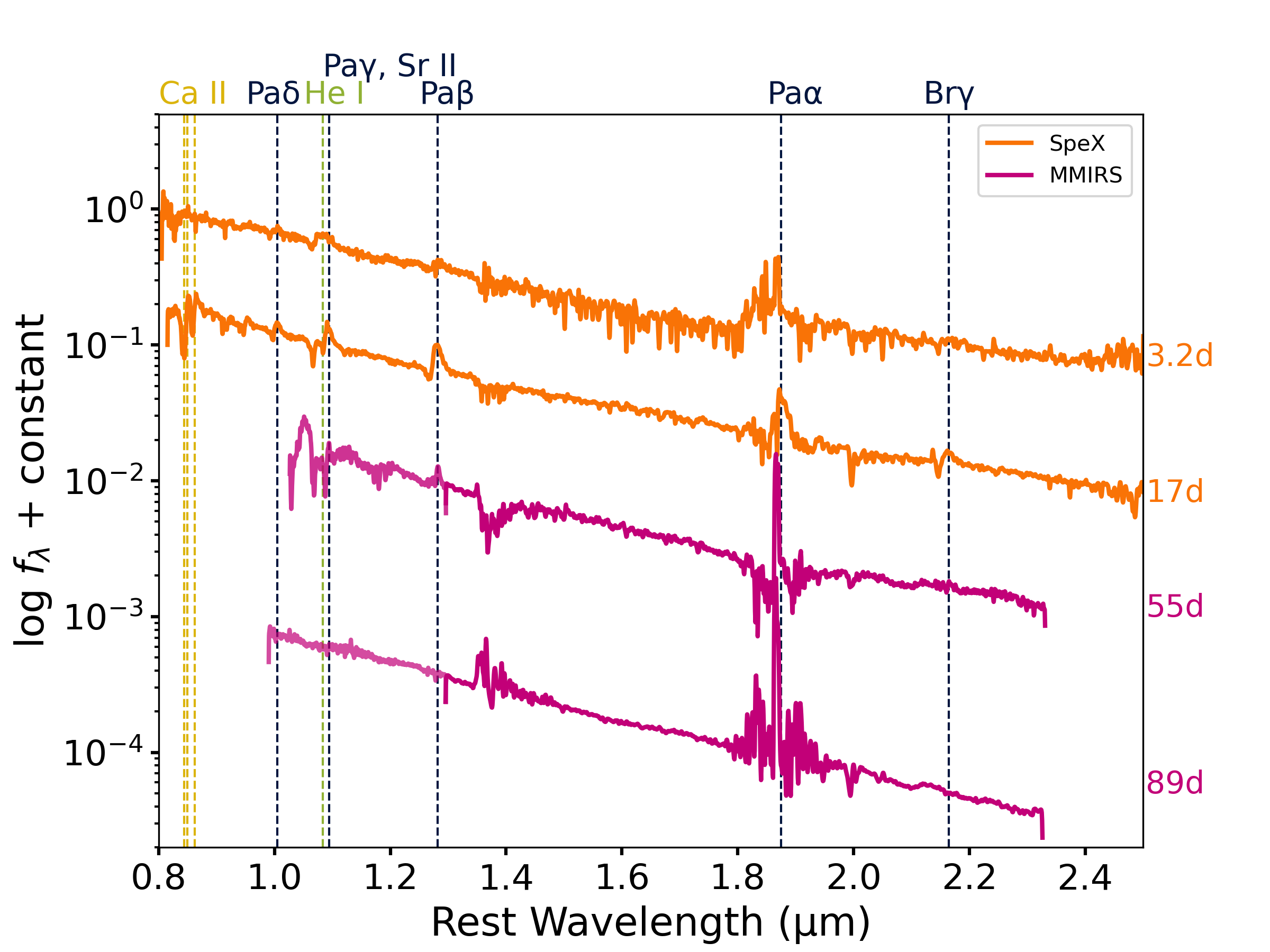}
    \caption{NIR spectral evolution of SN 2024abfl as described in Table \ref{tab:nirspec}. Notable lines are identified, and the dates are with respect to our assumed explosion epoch of MJD 60627.91.}
    \label{fig:NIRspec}
\end{figure}

Figure \ref{fig:NIRspec} shows the NIR spectral evolution from 3.2 to 89 days after the explosion.  The day 3.2 spectrum is fairly featureless, except for weak Paschen lines and possible \ion{He}{1} 10830  and \ion{Sr}{2} 10915 (blended with Pa$\gamma$). Possible absorption from \ion{C}{1} 10691 can also be seen. By day 17, the Paschen lines strengthen with corresponding P-Cygni absorption, and even Br$\gamma$ emission is seen. As the SN evolves along the plateau, these lines narrow and weaken, so by day 89 only Pa$\alpha$ is  detected in the noise of the telluric band. 

Unfortunately, due to their low luminosity and low occurrence rate, not many NIR spectra of LLSNe exist in the literature. For what NIR spectra are published we do see similar features as seen in SN~2024abfl. For instance, SN~2018is at 16.3 days looks similar to our 17 day spectrum of 2024abfl \citep{2025A&A...694A.260D}, while the day 55 spectrum shows similarities between the day 62 NIR spectra of SN~2005cs \citep{2009MNRAS.394.2266P} and the day 45 spectrum of SN~1997D \citep{2001MNRAS.322..361B}. 

\subsection{Late-time Nebular Spectra}
\label{sec:Nebular}
We obtained nebular spectra of SN 2024abfl at 319, 340, and 348 days post-explosion. In Figure \ref{fig:nebularcomp}, the nebular spectra of SN 2024abfl at 319 and 340 days are compared to similar epochs of SN 2005cs \citep{2009MNRAS.394.2266P}, which has a confirmed low mass RSG progenitor, SN 2018zd \citep{2021NatAs...5..903H}, which has been suggested as a possible ECSN (but was also proposed to be a low-mass Fe core collapse too \citep{2021arXiv210912943C}) in the same galaxy as SN~2024abfl, and SN 2018lab \citep{2023ApJ...945..107P}, a similar, well-studied LLSNe. SNe 2005cs, 2018lab, and 2024abfl all look nearly identical at this stage, whereas SN~2018zd displays less prominent Ca, O, and C lines.

The spectrum of SN 2024abfl at $\sim$300 days post-explosion resembles those of other faint IIP SNe at this stage, with prominent H$\alpha$, NaID, and the \ion{Ca}{2} triplet still visible and prominent. However, forbidden nebular lines are now among the most dominant features in the spectrum. The [\ion{Ca}{2}] $\lambda\lambda 7291,7324$ doublet is particularly prominent and the [\ion{O}{2}] $\lambda\lambda6300,6364$ doublet is clearly detected, though less visible.

This spectral morphology is characteristic of LLSNe and has been observed in well-studied events such as SN~2005cs \citep{2009MNRAS.394.2266P}, SN~2016bkv \citep{2018ApJ...861...63H}, and SN~2018lab \citep{2023ApJ...945..107P}. The prominence of [\ion{Ca}{2}] relative to [\ion{O}{1}] is commonly interpreted as evidence for a low-mass progenitor and a small oxygen-rich core, consistent with low explosion energies and modest $^{56}$Ni yields.

\begin{figure*}
    \centering
    \includegraphics[width=\linewidth]{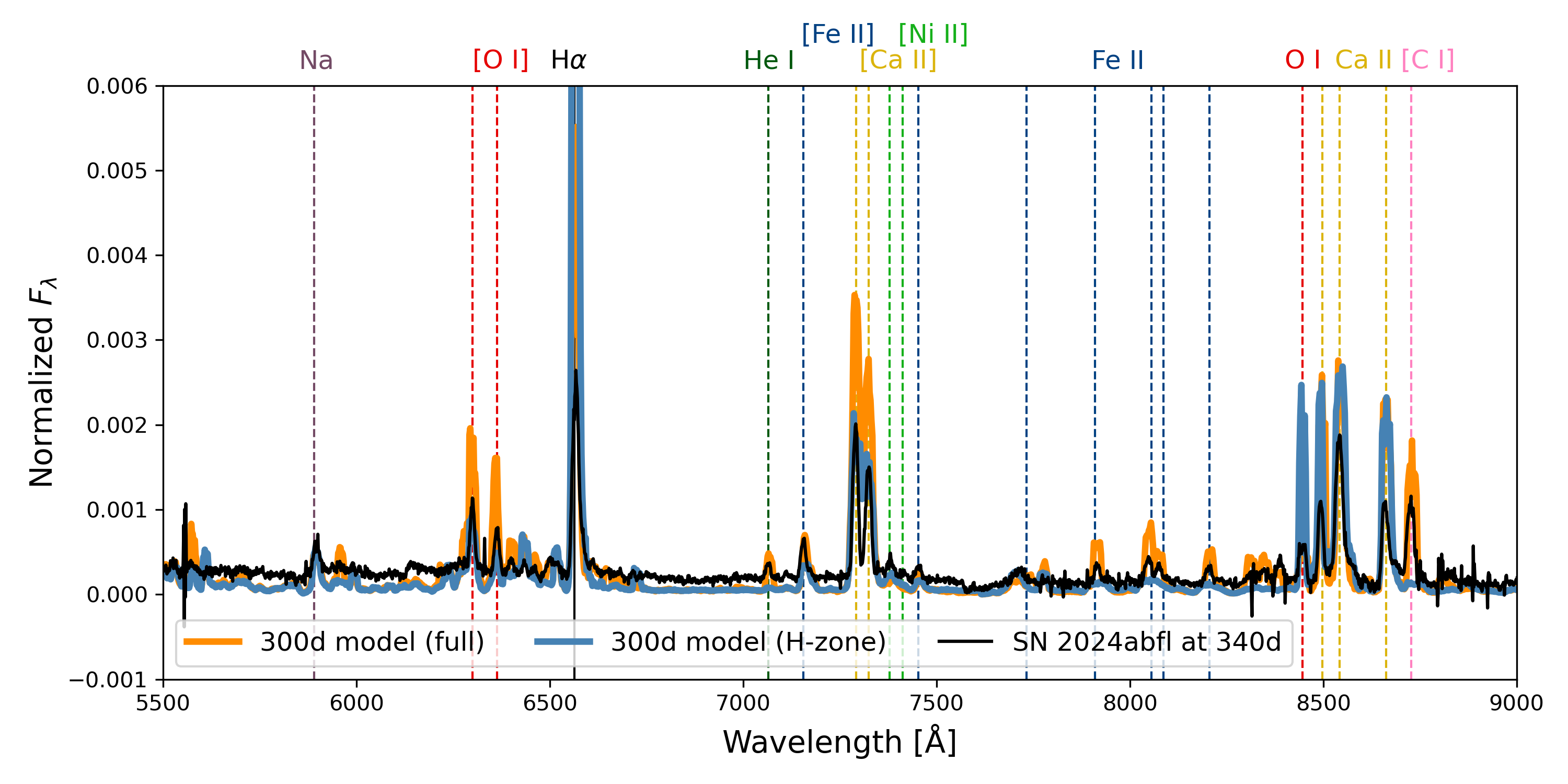}
    \caption{Late-time spectrum of SN 2024abfl taken 340 days post explosion compared to the $9~M_\odot$ \citet{2018MNRAS.475..277J} models. Both the models and the spectrum are normalized to the total flux over the wavelength range of the observed spectrum to highlight line ratio differences. The full model, orange, is the expected spectrum for an iron CCSN. The hydrogen-zone model, blue, should be similar to the nebular spectrum expected of an ECSN. The [\ion{C}{1}] $\lambda 8727$ line, absent in the hydrogen-zone model, is clearly present in SN 2024abfl.}
    \label{fig:jerkstrandcomp}
\end{figure*}

In Figure \ref{fig:jerkstrandcomp}, the nebular spectra of SN 2024abfl at 340 days post explosion is compared to the 300-day nebular models of a $9~M_\odot$ RSG progenitor presented in \citet{2018MNRAS.475..277J}. This mass is consistent with the ejecta mass derived from our \texttt{SNEC} modeling. The observed spectrum and both model spectra are normalized to the total flux over the wavelength range of the observed spectrum. The “pure hydrogen zone" model presented in \citet{2018MNRAS.475..277J} shows the signatures of a progenitor made up of only material from the hydrogen envelope. While the H-zone model is not an electron-capture model, an ECSN is expected to resemble this model. The most notable difference between the full Fe core-collapse model and the H-zone model is the lack of \ion{He}{1} $\lambda 7065$, \ion{Fe}{1} $\lambda 7900$--$8500$, and [\ion{C}{1}] $\lambda8727$ in the H-zone model.

Notably, SN 2024abfl displays a strong [\ion{C}{1}] $\lambda 8727$ feature, along with clear \ion{He}{1} $\lambda 7065$ and several \ion{Fe}{2} lines between $7900$--$8400$ \AA. Although weaker than in the full Jerkstrand model, these lines suggest the existence of He and O zones in the progenitor star at the time of collapse. This stellar composition indicates that SN 2024abfl is likely to be the result of iron core-collapse in a low-mass RSG. Taken together with the progenitor constraints from archival HST imaging (Section~\ref{sec:hst}) and the hydrodynamical modeling (Secion~\ref{sec:hydro}), the nebular spectra reinforce a coherent picture of SN 2024abfl arising from a $\sim9$--$12 M_\odot$ RSG undergoing iron core-collapse.

\section{Summary and Conclusions}
\label{sec:summary}
We have presented photometric and spectroscopic observations of SN 2024abfl, one of the faintest Type II SNe ever observed and one of the most comprehensively followed LLSNe to date. 
Its plateau, lasting $\sim$126 days with minimal decline, and slow color evolution place it firmly within the class of LLSNe. The rapid follow-up photometry and high-cadence early observations help tightly constrain the explosion epoch.  Its photospheric velocity evolution is low, never exceeding $\sim$3500 km s$^{-1}$, consistent with a low-energy explosion and a low ejecta mass, typical of other LLSNe such as SN 2005cs and SN 2018lab. The late-time luminosity implies a small synthesized $^{56}$Ni mass (of order a few $\times10^{-2} M_\odot$), again consistent with weak explosions of low-mass red supergiants.

The early spectra and light curve of SN 2024abfl reveal short-lived CSM interaction. Its fast rise, which cannot be reproduced by standard analytic shock-cooling models \citep{2017ApJ...838..130S}, is likely powered by this excess luminosity from ejecta-CSM interaction. Additionally, flash spectroscopy obtained within the first three days reveals broad emission lines that fade rapidly, signaling the presence of a confined, low-mass CSM. In the late stages of the plateau phase, asymmetric H$\alpha$ may also suggest interaction with a disc-like CSM at roughly 2 $\times$ 10$^{15}$ cm away. The fast rise, ledge features, H$\alpha$ profile, and failure of CSM-free models to match the early light curve indicate that SN~2024abfl belongs to an increasing sample of LLSNe with observed pre-explosion mass loss, emphasizing the importance of early-time observations.

At late times, the nebular spectra of SN 2024abfl display the expected features of an Fe-core-collapse event, consistent with a low-energy explosion from a low-mass red supergiant progenitor. Although some SNe in this luminosity regime, such as SN 2018zd and SN 2016bkv, have been proposed as ECSNe candidates, the photometric, spectroscopic, and nickel mass properties of SN 2024abfl more closely resemble a weak Fe-core collapse. 

SN 2024abfl reinforces the growing picture that low-luminosity SNe II span a continuum of explosion energies and progenitor masses. 
The detection of both ledge features and a rapid rise in luminosity demonstrates that confined CSM interaction is not exclusive to the most luminous SNe II, but may also be a common feature of low-mass RSG explosions. Continued early- and late-time observations of LLSNe are critical for uncovering the diversity of their pre-explosion mass-loss histories, mapping the range of explosion energies, and distinguishing between Fe-core-collapse and electron-capture channels. 
SN 2024abfl adds to the small but growing set of well-studied LLSNe and highlights the importance of rapid-response spectroscopy and deep nebular follow-up in understanding the final stages of low-mass stellar evolution.

\section*{Acknowledgments}
This work was supported by the NSF REU grant AST-2349202 and by the Nantucket Maria Mitchell Association. Based on observations obtained as part of (GN-2024B-LP-112; P.I. Sand $\&$ Andrews) at the international Gemini Observatory, a program of NSF NOIRLab, which is managed by the Association of Universities for Research in Astronomy (AURA) under a cooperative agreement with the U.S. National Science Foundation on behalf of the Gemini Observatory partnership: the U.S. National Science Foundation (United States), National Research Council (Canada), Agencia Nacional de Investigaci\'{o}n y Desarrollo (Chile), Ministerio de Ciencia, Tecnolog\'{i}a e Innovaci\'{o}n (Argentina), Minist\'{e}rio da Ci\^{e}ncia, Tecnologia, Inova\c{c}\~{o}es e Comunica\c{c}\~{o}es (Brazil), and Korea Astronomy and Space Science Institute (Republic of Korea). These observations were processed using DRAGONS (Data Reduction for Astronomy from Gemini Observatory North and South). This work was enabled by observations made from the Gemini North telescope, located within the Maunakea Science Reserve and adjacent to the summit of Maunakea. We are grateful for the privilege of observing the Universe from a place that is unique in both its astronomical quality and its cultural significance. 

Time-domain research by the University of Arizona team and D.J.S. is supported by National Science Foundation (NSF) grants 2108032, 2308181, 2407566, and 2432036 and the Heising-Simons Foundation under grant \#2020-1864. This work makes use of observations from the Las Cumbres Observatory network.  The LCO team is supported by NSF grant AST-230811.

The LBT is an international collaboration among institutions in the United States, Italy and Germany. LBT Corporation Members are: The University of Arizona on behalf of the Arizona Board of Regents; Istituto Nazionale di Astrofisica, Italy; LBT Beteiligungsgesellschaft, Germany, representing the Max-Planck Society, The Leibniz Institute for Astrophysics Potsdam, and Heidelberg University; The Ohio State University, and The Research Corporation, on behalf of The University of Notre Dame, University of Minnesota and University of Virginia.
This paper made use of the modsCCDRed data reduction code developed in part with funds provided by NSF Grants AST-9987045 and AST-1108693. 

Some observations reported here were obtained at the MMT Observatory, a joint facility of the University of Arizona and the Smithsonian Institution. 

\medskip
{\large \textit{Facilities:}} ADS, ATLAS, Boller $\&$ Chivens (B$\&$C), Gemini Observatory (Gemini North GMOS, GNIRS), Las Cumbres Observatory (QHY600, Sinistro, FLOYDS), LBT (MODS), MMT (BCS, Binospec, MMIRS), Swift (UVOT), WISeREP, ZTF

\medskip
{\large \textit{Software:}} astropy \citep{2013A&A...558A..33A, 2018AJ....156..123A}, corner \citep{2016JOSS....1...24F}, \texttt{DRAGONS} \citep{2023RNAAS...7..214L}, emcee \citep{2013PASP..125..306F}, Floyds pipeline \citep{2014MNRAS.438L.101V}, IRAF \citep{1986SPIE..627..733T}, \texttt{lcogtsnpipe} \citep{2016MNRAS.459.3939V}, Light Curve Fitting \citep{2023zndo...7872772H}, MatPLOTLIB \citep{2020Natur.585..357H}, NumPy \citep{2007CSE.....9...90H}, \texttt{SNEC} \citep{2015ApJ...814...63M}

\appendix

Table~\ref{tab:optspec} shows the complete optical spectroscopy log, and Table~\ref{tab:nirspec} shows the complete NIR spectroscopy log. All data will be made available as data behind the figure and are available on WISeRep (\url{https://www.wiserep.org})

\begin{deluxetable*}{lcccccc}
\tablecaption{Optical Spectroscopy of SN~2024abfl \label{tab:optspec}}
\tablehead{ \colhead{UT Date}    &\colhead{MJD}& \colhead{Phase}    &\colhead{Telescope+}   & \colhead{R}&  \colhead{Exposure Time}  \\[-6pt]
   \colhead{(y-m-d)}    &\colhead{} & \colhead{(days)} & \colhead{Instrument}  &\colhead{$\lambda$/$\Delta\lambda$}   & \colhead{(s)}   \  }
\startdata
2024-11-16 & 60630.49 & 2.6 & Gemini-N+GMOS & 1500 & 600 $\times$ 4 \\
2024-11-17 & 60631.48 & 3.6 & Gemini-N+GMOS & 1500 & 600 $\times$ 4 \\
2024-11-19 & 60633.16 & 5.2 & Bok+B\&C & 700 & 900 $\times$ 3 \\
2024-11-20 & 60634.15 & 6.2 & LBT+MODS & 2000 & 900 $\times$ 2 \\
2024-11-20 & 60634.21 & 6.3 & Bok+B\&C & 700 & 900 $\times$ 3 \\
2024-11-20 & 60634.48 & 6.6 & Gemini-N+GMOS & 1500 & 600 $\times$ 4 \\
2024-11-21 & 60635.37 & 7.4 & Bok+B\&C & 700 & 900 $\times$ 3 \\
2024-11-22 & 60636.44 & 8.5 & FTN+FLOYDS & 380 & 3600 \\
2024-11-23 & 60637.49 & 9.6 & MMT+BINO & 1300 & 900 $\times$ 3 \\
2024-11-25 & 60639.47 & 11.6 & FTN+FLOYDS & 380 & 3600 \\
2024-11-29 & 60643.14 & 15.2 & LBT+MODS & 2000 & 900 $\times$ 2 \\
2024-11-30 & 60644.48 & 16.6 & Gemini-N+GMOS & 1500 & 600 $\times$ 4 \\
2024-12-05 & 60649.44 & 21.5 & FTN+FLOYDS & 380 & 3600 \\
2024-12-07 & 60651.38 & 23.5 & MMT+BINO & 1300 & 900 $\times$ 3 \\
2024-12-12 & 60656.12 & 28.2 & LBT+MODS & 2000 & 900 $\times$ 2 \\
2024-12-14 & 60658.31 & 30.4 & LBT+MODS & 2000 & 900 $\times$ 2 \\
2024-12-16 & 60660.40 & 32.5 & LBT+MODS & 2000 & 900 $\times$ 2 \\
2024-12-26 & 60670.35 & 42.4 & MMT+BINO & 1300 & 900 $\times$ 2 \\
2024-12-31 & 60675.32 & 47.4 & Bok+B\&C & 700 & 1500 $\times$ 4 \\
2025-01-02 & 60677.73 & 49.3 & Bok+B\&C & 700 & 900 $\times$ 4 \\
2025-01-12 & 60687.32 & 59.4 & FTN+FLOYDS & 380 & 3600 \\
2025-01-20 & 60695.30 & 67.4 & FTN+FLOYDS & 380 & 3600 \\
2025-02-01 & 60707.64 & 79.2 & Bok+B\&C & 700 & 1500 $\times$ 3 \\
2025-02-03 & 60709.21 & 81.3 & Bok+B\&C & 700 & 900 $\times$ 4\\
2025-02-03 & 60709.33 & 81.4 & LBT+MODS & 2000 & 900 $\times$ 4\\
2025-02-04 & 60710.14 & 82.2 & MMT+BINO & 1300 & 900 $\times$ 3 \\
2025-02-05 & 60711.28 & 83.4 & FTN+FLOYDS & 380 & 3600 \\
2025-02-17 & 60723.32 & 95.4 & FTN+FLOYDS & 380 & 3600 \\
2025-02-20 & 60726.21 & 98.3 & MMT+BCH & 5800& 180 $\times$ 3 \\
2025-02-22 & 60728.19 & 100.3 & MMT+BCH & 800 & 180 $\times$ 3\\
2025-03-01 & 60735.27 & 107.4 & FTN+FLOYDS & 380 & 3600 \\
2025-03-17 & 60751.23 & 123.3 & FTN+FLOYDS & 380 & 3600 \\
2025-03-30 & 60764.19 & 136.3 & MMT+BINO & 1300 & 900 $\times$ 3 \\
2025-03-30 & 60764.23 & 136.3 & Gemini-N+GMOS & 1900 & 300 $\times$ 4 \\
2025-09-29 & 60947.57 & 319.7 & Gemini-N+GMOS & 1900 & 1200 $\times$ 4 \\
2025-10-20 & 60968.37 & 340.5 & MMT+BINO & 1300 & 1200 $\times$ 3 \\
2025-10-28 & 60976.33 & 348.4 & MMT+BCH & 1200  & 1800 $\times$ 4 \\
\hline
\enddata
\tablecomments{Phases are with respect to an estimated explosion epoch of MJD 60627.91.}
\end{deluxetable*}

\clearpage

\startlongtable
\begin{deluxetable*}{lccccc}
\tablecaption{NIR Spectroscopy of SN~2024abfl \label{tab:nirspec}}
\tablehead{ \colhead{UT Date}    &\colhead{MJD}& \colhead{Phase}    &\colhead{Telescope+}   &  \colhead{Exposure Time}  \\[-6pt]
   \colhead{(y-m-d)}    &\colhead{} & \colhead{(days)} & \colhead{Instrument}     &   \colhead{(s)}  }
\startdata
2024-11-17 & 60631.47 & 3.56  &  GN+GNIRS  &  180 $\times$ 2 \\
2024-11-30 & 60644.47 & 16.56  &  GN+GNIRS  & 180 $\times$ 2 \\
2025-01-08 & 60683.23 & 55.31 & MMT+MMIRS & 750 $\times$ 2  \\
2025-02-11 & 60717.31 & 89.40 & MMT+MMIRS &  1100 $\times$ 2 \\
\hline
\enddata
 \tablecomments{Phases are with respect to an estimated explosion epoch of MJD 60627.91. Exposure numbers include a full ABBA sequence.}
 \end{deluxetable*}

\bibliography{SN2024abfl}{}
\bibliographystyle{aasjournalv7}
\end{document}